\documentclass[sigconf,nonacm]{acmart}
\usepackage{multirow}
\usepackage{algorithm}
\usepackage{algpseudocode}
\usepackage{latexsym}
\usepackage{amsmath}
\usepackage[inline]{enumitem}
\usepackage{float}
\usepackage{xcolor} 
\usepackage{xspace}
\usepackage{subcaption}
\usepackage{graphicx} 
\usepackage{adjustbox} 
\usepackage{booktabs} 
\usepackage{longtable} 
\usepackage{xcolor}
\usepackage{soul} 
\usepackage{tabularx,multirow,booktabs,blindtext} 
\usepackage{graphicx}
\usepackage{listings, mdframed}
\usepackage{xcolor}  
\usepackage{amsmath}
\usepackage{algpseudocode}
\usepackage{algorithm}
\usepackage{enumitem} 
\usepackage{colortbl}
\usepackage{hhline} 
\usepackage{highlight}

\newcommand{\name}{LLMCloudHunter\xspace}
\newcommand{\sig}{\textit{Sigma}\xspace}
\newcommand{\model}{GPT-4o\xspace}
\newcommand{\noimagesmodel}{BlindHunter\xspace}
\newcommand{\noapimodel}{NoAPIHunter\xspace}
\newcommand{\nooptimizationmodel}{UnoptimizedHunter\xspace}

\definecolor{ProcessBlue}{RGB}{0, 136, 206}

\lstdefinestyle{yaml}{
    backgroundcolor=\color{white},
    basicstyle=\ttfamily\footnotesize,
    comment=[l]{:},
    commentstyle=\color{black}
 }

\renewcommand{\opacity}{50}

\AtBeginDocument{%
  }

\setcopyright{acmlicensed}
\copyrightyear{2024}
\acmYear{2024}
\acmDOI{XXXXXXX.XXXXXXX}

\acmConference[ACSAC '24]{Make sure to enter the correct conference title from your rights confirmation emai}
{December 09--13, 2024}{Waikiki, Hawaii, USA}
\acmBooktitle{Waikiki '24: Annual Computer Security Applications Conference,
 December 09--13, 2024, Waikiki, Hawaii, USA} 
\acmISBN{978-1-4503-XXXX-X/24/12}

\begin{document}

\title{\name: Harnessing LLMs for Automated Extraction of Detection Rules from Cloud-Based CTI}

\author{Yuval Schwartz*, Lavi Benshimol*, Dudu Mimran, Yuval Elovici, Asaf Shabtai}
\thanks{*Both authors contributed equally to this research.}
\affiliation{%
  \institution{Department of Software and Information Systems Engineering}
  \institution{Ben-Gurion University of the Negev}
  \country{}}

\renewcommand{\shortauthors}{Schwartz et al.}

\begin{abstract}
As the number and sophistication of cyber attacks have increased, threat hunting has become a critical aspect of active security, enabling proactive detection and mitigation of threats before they cause significant harm.
Open-source cyber threat intelligence (OSCTI) is a valuable resource for threat hunters, however, it often comes in unstructured formats that require further manual analysis.
Previous studies aimed at automating OSCTI analysis are limited since (1) they failed to provide actionable outputs, (2) they did not take advantage of images present in OSCTI sources, and (3) they focused on on-premises environments, overlooking the growing importance of cloud environments.
To address these gaps, we propose \name, a novel framework that leverages large language models (LLMs) to automatically generate generic-signature detection rule candidates from textual and visual OSCTI data.
We evaluated the quality of the rules generated by the proposed framework using 12 annotated real-world cloud threat reports.
The results show that our framework achieved a precision of 92\% and recall of 98\% for the task of accurately extracting API calls made by the threat actor and a precision of 99\% with a recall of 98\% for IoCs.
Additionally, 99.18\% of the generated detection rule candidates were successfully compiled and converted into \textit{Splunk} queries.
\end{abstract}

\ccsdesc[500]{Security and privacy}
\ccsdesc[500]{Computing methodologies~Natural language processing}
\ccsdesc[500]{Information systems~Information systems applications}

\keywords{Cyber threat intelligence (CTI), Large language model (LLM), Threat hunting, Cloud, Sigma rules}

\received{28 May 2024}
\received[revised]{7 August 2024}
\received[accepted]{19 August 2024}

\maketitle
\section{\label{sec:intro}Introduction}

The rapid evolution of technology, digitization, and application development has been accompanied by an increase in the number of cyberattacks~\cite{ozkan2024comprehensive}, raising concerns about the security risks associated with these advancements.
In the face of these concerns, organizations have adopted dynamic defensive strategies in addition to the traditional reactive measures employed~\cite{nour2023survey}.
One such strategy is threat hunting, a proactive approach aimed at searching for and mitigating undetected threats in a network or system~\cite{what_is_threat_hunting}.
Security operation centers' threat hunters try to minimize the damage caused by threat actors by shortening the time window between intrusion and discovery~\cite{arafune2022design}.
In their comprehensive survey, \citeauthor{nour2023survey}~\cite{nour2023survey} stated that the threat hunting methodology consists of three main principles: (1) formulating and testing hypotheses about the threat actor and their actions; (2) utilizing existing information for an intelligence-driven investigation; and
(3) leveraging data analysis techniques and machine learning algorithms to effectively handle vast amounts of data.

The second principle involves collecting and analyzing publicly available information about potential and active threats from blogs, forums, and other digital sources.
Open-source cyber threat intelligence (OSCTI) is one of the most commonly used sources of information among security personnel according to the SANS 2023 CTI survey~\cite{sans_2023_cti_survey}.
However, various challenges arise when using OSCTI.
The first and main challenge is that OSCTI often comes in non-uniform and unstructured formats, such as text and images, rather than more actionable information/data (e.g., detection rules)~\cite{rahman2020literature}.
As a result, manual analysis by human experts is required to derive meaningful and actionable insights~\cite{rahman2023attackers}.
Another challenge is the increasing amount of available information (i.e., CTIs), necessitating the automation of OSCTI analysis~\cite{ozkan2024comprehensive}.

Previous studies on threat hunting introduced various methodologies, some of which incorporated natural language processing (NLP) techniques, to automate the extraction and enrichment of information from OSCTI textual data. 
However, the methods presented in these studies suffer from three main limitations: (1) they provide structured but limited insights, such as identified entities and their relationships or attack techniques, necessitating further processing to generate actionable outputs; an exception is the approach presented by \citeauthor{gao2021enabling}~\cite{gao2021enabling}, in which the authors developed proprietary, non-standard graph-based queries using static rules (regexes) that require substantial customization for application with standard tools and on-premises environments; (2) these studies, including the work of \citeauthor{gao2021enabling}, do not take advantage of visual components, such as images, which may be present in OSCTI data; and (3) many of these methodologies were primarily developed for on-premise environments, limiting their effectiveness and relevance in cloud-centric environments.

Cloud computing has become an integral component in the modern enterprise landscape, valued for its benefits which include scalability, cost-effectiveness, and flexibility~\cite{ramchand2021enterprise}.
It employs a shared responsibility model for security, in which both the provider and the consumer play roles in securing cloud-resident infrastructure and cloud-delivered applications~\cite{ahmadi2024systematic}.
This model presents unique challenges in threat hunting, as traditional security methodologies often fall short in addressing the dynamic and distributed nature of cloud environments~\cite{tabrizchi2020survey}.
Among these challenges is the fact that in some cloud technologies (e.g., serverless), access to data for threat hunting is limited to application-level logs (APIs, storage access, etc.), and important infrastructure- (system)-level data (e.g., virtual machines and network) can only be accessed by the cloud provider~\cite{tamburri2020cloud}.
This is exacerbated by the fact that the exploitation of cloud-based threat intelligence has not yet reached maturity. 
The work of \citeauthor{fengrui2024few}~\cite{fengrui2024few} is the only study that extends beyond on-premise OSCTI, however rather than providing actionable output, their framework extracts MITRE ATT\&CK tactics, techniques, and procedures (TTPs)~\cite{mitre_attack_ttp}.
These gaps highlight the need for innovative OSCTI analysis approaches suited to the unique security challenges of cloud environments; such challenges can be addressed by integrating OSCTI analysis results within practical, actionable security measures~\cite{kaur2023artificial}.

In this paper, we present \name, a novel framework that leverages pretrained large language models (LLMs) to generate detection rule candidates from unstructured OSCTIs automatically.
\name generates \sig rule~\cite{sigma_rule} candidates from both textual and visual cyber threat information, using an innovative, automated data extraction and processing framework that leverages LLMs and employs various techniques to address their limitations (e.g., unstructured output and hallucinations).

\sig rules, provided in a generic and open signature format written in YAML, enable the creation and sharing of detection methods across security information and event management (SIEM) systems. 
Fig.~\ref{fig:clout_overview} presents our LLM pipeline for \sig candidate generation; as can be seen, textual and visual OSCTI data is processed first, converting it into semi-structured paragraphs in the preprocessing phase. 
It then extracts API calls (that are unique entities to threat hunting in cloud environments) and MITRE ATT\&CK TTPs from the paragraphs and generates initial \sig candidates (in the Paragraph-Level phase).
Finally, it consolidates the candidates from all paragraphs, verifies their syntactic and logical correctness, eliminates duplication, and enriches them with identified indicators of compromise (IoCs) (in the OSCTI-Level phase).
An example of a \sig rule generated by \name is illustrated in Listing~\ref{lst:sigma_candidate_example}, with a demonstration of its generation process in Appendix~\ref{app:example}.
 
We evaluated the efficacy and precision of the \sig candidates generated using 12 cloud-related OSCTI sources that we identified.
The evaluation was performed using common entity and relationship extraction metrics, and the results were validated against a ground truth carefully defined by our research team.
Additionally, we introduced a set of criteria specifically designed to test each \sig candidate's functionality in the operational context of OSCTI.
This comprehensive evaluation ensures that the rules generated not only meet syntactic standards but are also operationally effective in addressing the dynamic and complex nature of cloud-based cyber threats.
We also conducted an ablation study, systematically removing components of the framework to pinpoint their individual contributions to \name's overall efficacy.
The results show that our framework achieved a precision of 92\% and recall of 98\% for the task of accurately extracting threat actors' API calls, and a precision of 99\% with a recall of 98\% for IoCs.
Moreover, 99.18\% of the generated \sig candidates were successfully converted into \textit{Splunk} queries.
In terms of overall performance, i.e., including the extraction of API calls, IoCs, MITRE ATT\&CK TTPs, and request parameters, our framework achieved 80\% and 83\% precision and recall, respectively.

To summarize, the main contributions of this paper are:
\begin{itemize}[nosep,leftmargin=*]
  \item[1.] A novel LLM-based framework for the automatic generation of \sig candidates from unstructured OSCTI, which integrates both textual and visual information.
  While our framework focuses on cloud environments, it can be adapted for use with on-premise-related CTI. 
  \name utilizes a pretrained LLM, thus providing flexibility in modifying the framework and updating the underlying LLM, and does not require "heavy" model training.
  \item[2.] An annotated dataset consisting of 12 cloud-related OSCTI posts, complete with entities and their relationships, as well as \sig rules.
  This dataset supports the training and evaluation of our framework.
  \item[3.] Insights on the application of LLMs for complex NLP tasks in the field of cybersecurity, pertaining to prompt engineering techniques and the effective use of models' features and parameters.
  \item[4.] A comprehensive evaluation that assesses the accuracy and correctness of the \sig candidates generated.
  \item[5.] We make both our code and cloud CTI dataset available to the research community on GitHub\footnote{To preserve anonymity, the code and dataset will be available upon paper acceptance.}.
\end{itemize}

\begin{lstlisting}[
    style=yaml,
    keywordstyle=\color{ProcessBlue},
    frame=single,
    keywords={title, id, status, description, references, author, date, modified, tags, logsource, product, service, detection, selection_event, eventSource, eventName, requestParameters, key, selection_ip_address, sourceIPAddress, condition, falsepositives, level},
    literate={'}{{\textquotesingle}}1,
    caption=A Sigma rule generated by \name.,
    label=lst:sigma_candidate_example,
    captionpos=b
    ]
title: Access to Terraform File from Malicious IPs
description: Detects requests for terraform.tfstate file
  from known malicious IPs. This file contains sensitive
  infrastructure information and secrets, indicating
  potential compromise or unauthorized access.
references:
    - https://sysdig.com/blog/cloud-breach-terraform-data-
  theft/
    - https://docs.aws.amazon.com/AmazonS3/latest/API/
  API_GetObject.html
author: LLMCloudHunter
tags:
    - attack.collection
    - attack.t1530
logsource:
    product: aws
    service: cloudtrail
detection:
    selection_event:
        eventSource: s3.amazonaws.com
        eventName: GetObject
        requestParameters.key: terraform.tfstate
    selection_ip_address:
        sourceIPAddress:
            - 80.239.140.66
            - 45.9.148.221
            - 45.9.148.121
            - 45.9.249.58
    condition: selection_event and selection_ip_address
falsepositives:
  - Automated CI/CD pipeline operations
  - DevOps engineers manually running Terraform commands
level: high
\end{lstlisting}

\section{\label{sec:related}Related Work}
In this section, we provide an overview of recent studies focused on unstructured OSCTI analysis (also summarized in Table~\ref{tab:related_work}).

\noindent\textbf{OSCTI Analysis Techniques.}
The development of efficient threat hunting mechanisms that leverage OSCTI has resulted in a wide range of research methodologies, each using different approaches to analyze and interpret OSCTI data.
Within each OSCTI, key information (e.g., IoC or TTPs) is often implicit and requires the use of a different extraction approach. 

NLP techniques have been utilized extensively for OSCTI analysis in methods including: Casie~\cite{satyapanich2020casie}, Extractor~\cite{satvat2021extractor}, Open-CyKG~\cite{sarhan2021open}, SecIE~\cite{park2022full}, and CyberEntRel~\cite{ahmed2024cyberentrel}.
These methods leveraged advanced NLP models (e.g., BiLSTM, BERT, RoBERTa) to extract actionable insights from OSCTI text. 
However, to adapt these models to the cyber threat domain, a significant amount of preprocessing and fine-tuning is required.
TTPDrill~\cite{husari2017ttpdrill} and THREATRAPTOR~\cite{gao2021enabling} implement an unsupervised NLP pipeline that employs rule-based and information retrieval techniques.
While this approach reduces the need for extensive model training, it is not flexible, and significant customization is needed for use in cloud environments.
This is due to fundamental differences in terminology and data types between traditional on-premise environments and cloud environments, as well as the dynamic nature of cloud architectures, which continuously evolve with new services and configurations.

The introduction of LLMs has led to a paradigm shift in OSCTI processing, with research demonstrating their ability to extract meaningful and structured data from OSCTI text.
Utilizing GPT-3.5, \citeauthor{purba2023extracting}~\cite{purba2023extracting} and \citeauthor{siracusano2023time}~\cite{siracusano2023time} addressed tasks ranging from the extraction of IoCs to the generation of structured CTI format (e.g., STIX), respectively, while \citeauthor{liu2023constructing}~\cite{liu2023constructing} applied ChatGPT to construct graphical representations of OSCTI data.
\citeauthor{hu4671345llm}~\cite{hu4671345llm} and \citeauthor{fengrui2024few}~\cite{fengrui2024few} expanded upon these capabilities by utilizing both pretrained and fine-tuned LLM models.
They employed GPT-3.5 and ChatGPT for data annotation and augmentation, respectively, to prepare datasets for fine-tuning the LLaMA2-7B model.
\citeauthor{hu4671345llm}~\cite{hu4671345llm} applied the fine-tuned LLaMA2-7B to construct knowledge graphs, while \citeauthor{fengrui2024few}~\cite{fengrui2024few} focused on TTP classification.

In this research, we are the first to develop an end-to-end framework based on a pretrained LLM, demonstrating the potential of LLMs in processing OSCTI and generating actionable \sig rules.
Moreover, our framework integrates visual analysis capabilities, expanding the scope of OSCTI analysis beyond previous text-centric methodologies.
By leveraging pretrained LLMs, we avoid the need for rule-based methods or training customized models with dedicated datasets.
Additionally, our framework focuses on generating rules for cloud environments, which has not been addressed in prior studies.

\textbf{Datasets.} In terms of OSCTI datasets, the study introducing TTPDrill~\cite{husari2017ttpdrill} used a dataset of semi-structured Symantec threat reports, from which threat actions were manually extracted.
Similarly, \citeauthor{satyapanich2020casie}~\cite{satyapanich2020casie} employed cybersecurity news articles published on CyberWire\footnote{https://thecyberwire.com/}, which were annotated before evaluation.
The study presenting Extractor~\cite{satvat2021extractor} used multiple structured OSCTI sources, including Microsoft, Symantec, Threat Encyclopedia, and Virus Radar.
Open-CyKG~\cite{sarhan2021open} used a structured OSCTI database focusing on malware.
ThreatRaptor~\cite{gao2021enabling} utilized the DARPA TC dataset, incorporating semi-structured OSCTIs, along with IoCs and relevant event log entries for each attack incident.
SecIE~\cite{park2022full} used 133 unstructured labeled threat reports from various threat intelligence vendors.
CyNer~\cite{fujii2022cyner} and TriCTI~\cite{liu2022tricti} developed a custom web crawler to retrieve unstructured OSCTIs across selected high-quality websites (e.g., Kaspersky, Symantec, and Fireye) and manually annotated a subset for evaluation purposes.
LLM-TIKG~\cite{hu4671345llm} also developed a custom web crawler to collect OSCTIs from selected platforms, but this study differs from the study presenting CyNer in that it utilizes an LLM (GPT) for annotation.
\citeauthor{liu2023constructing}~\cite{liu2023constructing} manually collected OSCTIs from public sites, and for each OSCTI, they selected the paragraphs that refer to the target technique to increase information density.
LADDER~\cite{alam2023looking} used OSCTI reports related to a specific set of malware, employing the BRAT~\cite{stenetorp2012brat} NLP method to annotate the concepts and their relationships in the text.
\citeauthor{purba2023extracting}~\cite{purba2023extracting} analyzed a dataset comprising 150 cyber threat related tweets. 
aCTIon~\cite{siracusano2023time} manually collected OSCTI posts and their respective STIX bundles and used expert-based annotation to create the ground truth.
CyberEntRel~\cite{ahmed2024cyberentrel} collected OSCTI reports from high-quality vendors (e.g., Microsoft, Cisco, McAfee, and Kaspersky) and performed keyword-based data extraction.
\citeauthor{fengrui2024few}~\cite{fengrui2024few} used the MITRE ATT\&CK dataset, structured in STIX 2.1 JSON format, which is a tagged and organized collection of adversary tactics and techniques.

In contrast to prior studies that primarily used semi-structured and on-premise-related datasets, we use 12 \emph{unstructured}, publicly available \emph{cloud-based} posts and reports sourced from various publishers.
These OSCTI reports, which describe AWS cloud incidents, were systematically annotated by our research team to develop a robust ground truth for development and evaluation.

\textbf{Extractions and Outputs.}
Previous studies produced a variety of outputs with different levels of utility and applicability.
\citeauthor{liu2022tricti}~\cite{liu2022tricti} and \citeauthor{purba2023extracting}~\cite{purba2023extracting} focused on extracting IoCs, while \citeauthor{fengrui2024few}~\cite{fengrui2024few} extracted TTPs. 
The studies presenting TTPDrill~\cite{husari2017ttpdrill}, Cyner~\cite{fujii2022cyner}, and aCTIon~\cite{siracusano2023time} converted unstructured OSCTIs into structured representations using the STIX format, which facilitates the systematic sharing and analysis of threat information.
A more advanced approach was used in Extractor~\cite{satyapanich2020casie} and ThreatRaptor~\cite{gao2021enabling}, in which threat behavior graphs are created; and in Casie~\cite{satyapanich2020casie}, Open-CyKG~\cite{sarhan2021open}, SecIE~\cite{park2022full}, LADDER~\cite{alam2023looking}, aCTIon~\cite{liu2023constructing}, LLM-TIKG~\cite{hu4671345llm}, CyberEntRel~\cite{ahmed2024cyberentrel}, in which knowledge graphs are generated.
Both approaches interrelate entities with associated actions and artifacts (i.e., IoCs and TTPs), providing structured insights into attack strategies through graph-based representations.
While the approaches highlighted above provide valuable contextual information, further processing is required to transform the representations into actionable defense mechanisms.

To address this, in their study, \citeauthor{gao2021enabling} presented a framework for converting OSCTI data into a threat behavior graph and associated domain-specific queries.
Both frameworks go beyond simply identifying and contextualizing threat data, by developing operational detection rules and queries.
The detection rule candidates generated by \name, however, are in the known open-source \sig structure.
This widely used generic signature format is inherently suitable for integration in various application environments and SIEMs.
By capturing the entities, relations, IoCs, and TTPs identified in OSCTI, \name translates threat intelligence into applicative \sig candidates.

\begin{table*}
    \begin{adjustbox}{width=0.999\textwidth,center}
        \begin{tabular}{lllllclccccc}
            \toprule
            \multirow{3}{*}{\textbf{Reference}} & \multirow{3}{*}{\textbf{Year}} & \multirow{3}{*}{\textbf{Technique}} & \multirow{3}{*}{\textbf{Dataset}} & \multirow{2}{*}{\textbf{Target}} & \multirow{2}{*}{\textbf{Image}} & \multirow{3}{*}{\textbf{Output}} & \multicolumn{5}{c}{\textbf{Extraction}} \\
            \cmidrule{8-12}
            & & & & \multirow{2}{*}{\textbf{Environment}} & \multirow{2}{*}{\textbf{Processing}} & & \multirow{2}{*}{\textbf{Entities}} & \multirow{2}{*}{\textbf{Relations}} & \multirow{2}{*}{\textbf{IoCs}} & \multirow{2}{*}{\textbf{TTPs}} & \textbf{Detection} \\
            & & & & & & & & & & & \textbf{Queries/Rules} \\
            \midrule

            TTPDrill~\cite{husari2017ttpdrill} & \citeyear{husari2017ttpdrill} & Unsupervised NLP & Symantec & On-premise & $\times$ & STIX & \checkmark & \checkmark & \checkmark & \checkmark & \\
    
            Casie~\cite{satyapanich2020casie} & \citeyear{satyapanich2020casie} & BiLSTM & CyberWire & On-premise & $\times$ & Knowledge Graph & \checkmark & \checkmark & & \checkmark & \\
    
            Extractor~\cite{satvat2021extractor} & \citeyear{satvat2021extractor} & BERT-BiLSTM & APT Repository, & On-premise & $\times$ & Threat Behavior Graph & \checkmark & \checkmark & \checkmark & & \\
            & & & Microsoft, Symantec, \\
            & & & Threat Encyclopedia, \\
            & & & Virus Radar \\
    
            Open-CyKG~\cite{sarhan2021open} & \citeyear{sarhan2021open} & BiLSTM & MalwareDB & On-premise & $\times$ &  Knowledge Graph & \checkmark & \checkmark & & \\
    
            ThreatRaptor~\cite{gao2021enabling} & \citeyear{gao2021enabling} & Unsupervised NLP & DARPA TC & On-premise & $\times$ & Threat Behavior Graph, &\checkmark &\checkmark &\checkmark &\checkmark &\checkmark \\
            & & & & & & TBQL Queries \\
    
            SecIE~\cite{park2022full} & \citeyear{park2022full} & BERT & CVE & On-premise & $\times$ & Knowledge Graph & \checkmark & \checkmark & \checkmark & & \\
    
            CyNER~\cite{fujii2022cyner} & \citeyear{fujii2022cyner} & BERT & Custom & On-premise & $\times$ & STIX & \checkmark & \checkmark & \checkmark & & \\
    
            TriCTI~\cite{liu2022tricti} & \citeyear{liu2022tricti} & BERT & Custom & On-premise & $\times$ & Labeled IoCs & & & \checkmark & \checkmark & \\
    
            LADDER~\cite{alam2023looking} & \citeyear{alam2023looking} & BERT & Custom & On-premise & $\times$ & Knowledge Graph & \checkmark & \checkmark & \checkmark & \checkmark & \\
    
            \citeauthor{purba2023extracting}~\cite{purba2023extracting} & \citeyear{purba2023extracting} & GPT-3.5 & Twitter Posts & On-premise & $\times$ & Labeled IoCs & \checkmark & & \checkmark \\
    
            aCTIon~\cite{siracusano2023time} & \citeyear{siracusano2023time} & GPT-3.5 & Custom & On-premise & $\times$ & STIX & \checkmark & \checkmark & \checkmark & \checkmark &\\

            \citeauthor{liu2023constructing}~\cite{liu2023constructing} & \citeyear{liu2023constructing} & ChatGPT & Custom & On-premise & $\times$ & Knowledge Graph & \checkmark & \checkmark & \checkmark  & & \\
    
            LLM-TIKG~\cite{hu4671345llm} & \citeyear{hu4671345llm} & Fine-tuned LLaMA-2-7B & Custom & On-premise & $\times$ & Knowledge Graph & \checkmark & \checkmark & \checkmark & \checkmark \\

            CyberEntRel~\cite{ahmed2024cyberentrel} & \citeyear{ahmed2024cyberentrel} & RoBERTa-BiGRU-CRF & Custom & On-premise & $\times$ & Knowledge Graph & \checkmark & \checkmark & & & \\
    
            \citeauthor{fengrui2024few}~\cite{fengrui2024few} & \citeyear{fengrui2024few} & Fine-tuned LLaMA-2-7B & ATT\&CK STIX Data & On-premise, & $\times$ & MITRE ATT\&CK TTPs & & & &\checkmark & \\
            & & & & Cloud & & \\
     
            \hline
            \textbf{Our Framework} & \textbf{2024} & \textbf{\model} & \textbf{Custom} & \textbf{Cloud} & \textbf{\checkmark} & \textbf{Sigma Rules} & \checkmark & \checkmark & \checkmark & \checkmark & \checkmark \\
            \bottomrule
        \end{tabular}
    \end{adjustbox}
    \caption{Comparison of studies utilizing OSCTI inputs.}
    \label{tab:related_work}
\end{table*}

\section{\label{sec:method}Proposed Method}
\begin{figure*}
  \includegraphics[width=\textwidth]{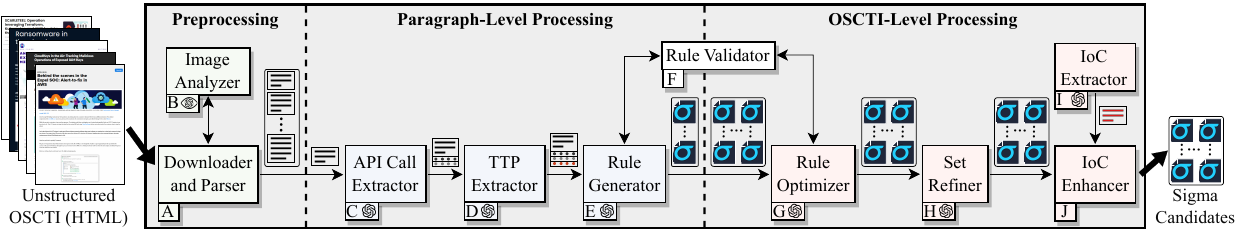}
  \caption{Overview of the \name framework.}
  \label{fig:clout_overview}
\end{figure*}

In this section, we describe our proposed framework, \name, and how it leverages OpenAI's \model~\cite{openai_models} model to process cloud-based OSCTIs and generate \sig candidates.
\name's pipeline (see Fig.~\ref{fig:clout_overview}) consists of three main phases: \emph{Preprocessing}, \emph{Paragraph-Level Processing}, and \emph{OSCTI-Level Processing}; these phases are described in the subsections that follow.

\noindent\textbf{Relevant Entities for Threat Hunting in Cloud Environments.}
The atomic units in cloud application logs are cloud API calls, which describe system and application activities that potentially provide traces of threat behavior.
An example of an API call may be the \textit{GetFunction} action, which requests information about a function.
Therefore, the information used to generate \sig candidates for threat hunting in cloud environments includes entities such as IP addresses and user agents, similar to on-premise environments, as well as API calls that are unique to cloud environments.

We differentiate between primary (essential) entities and contextual entities.
Primary entities are required for the correct execution of generated \sig candidates in SIEM systems.
A mistake in extracting a relationship that includes a primary entity will result in incorrect “hunting” activity.
Primary entities in cloud environments include API calls (e.g., \textit{GetFunction}) as well as the request parameters of that API call (e.g., \textit{requestParameters.functionName: respondUser}), IoCs (including IP addresses and user agents), log source (e.g., \textit{AWS CloudTrail}), and event source (e.g., \textit{lambda.amazonaws.com}).
Contextual entities do not impact the correctness of the detection rule logic; however, they provide additional contextual information to the threat hunter, making the investigation of a case more efficient.
Contextual entities include the title and description of the \sig rule, TTPs, false positives, and criticality level.

\subsection{OSCTI Preprocessing}\label{subsec:preprocess}
OSCTI varies in terms of the type and format, depending on the publishing platform, the author, the nature of the collected information, and its intended purpose.
Due to this lack of uniformity, preliminary steps must be performed to standardize the format.
Such steps enable the data to be automatically and effectively handled by subsequent processing components.

\noindent\textbf{Downloader and Parser.}
The automated OSCTI preprocessing phase begins by downloading the OSCTI HTML code, using a web scraping tool such as BeautifulSoup~\cite{beautiful_soup} (A in Fig.~\ref{fig:clout_overview}).
By examining the web page elements, \name pinpoints the beginning and end of the relevant content, excluding irrelevant elements (such as sidebars and advertisements).
In the next step, these HTML layout elements are converted into a unified text format, based on the following guidelines:
\begin{enumerate*}
    \item Preserve spacing to separate content types such as paragraphs and code sections, maintaining their original layout.
    \item Mark headings (h1, h2, etc.) to maintain the hierarchical structure of the original HTML content.
    \item Parse HTML code encompassing tables and nested lists to preserve their structural properties.
    For example, a tab character is employed in lists to signify nested items, whereas in tables, the `|' symbol is used to demarcate columns.
    \item Identify and embed image URLs as placeholders within the text, positioning them according to their original placement in the report.
\end{enumerate*}

After converting the HTML into a unified text format, we employed a targeted approach to exclude non-essential content (including headings, subheadings, and the corresponding paragraphs).
Such content is identified by indicative keywords that suggest repetitive and redundant information.
Examples of this type of content include overviews, recommendations, and concluding paragraphs.
For instance, if a `recommendations' paragraph appears under an h2 heading, we remove the paragraph and any subsequent content until the next h2 (or h1) heading is encountered, as recommendations are not part of the attack description and often include marketing content.
This approach effectively removes non-essential or duplicated content nested under the identified headings.
The filtered version of the output is then passed on to the next component in the framework. 
The full output, which includes all content, will be used in the \textit{OSCTI-Level Processing} phase.

\noindent\textbf{Image Analyzer.}
Continuing with the \textit{Preprocessing} phase, the \textit{Image Analyzer} (B in Fig.~\ref{fig:clout_overview}) handles the images extracted from the URL in the previous step.
Each image in the OSCTI is analyzed and converted into a textual representation. 
This conversion is performed by using the inherent visual analysis capabilities of \model.
The model is equipped with a tailored prompt designed specifically for cybersecurity analysis, as presented in Fig.~\ref{fig:image_analysis_prompt}.
Multiple images can be handled simultaneously to reduce the execution time and improve efficiency.
This component accurately transcribes the images' content into text.
Processing and utilizing these images is required since they often contain important information (e.g., API calls and IoCs) that appears only in the image and not in the text.
The information extracted from the images is then integrated back into the OSCTI formatted text in their original locations, preserving essential details and enriching the text for further analysis.

\subsection{Paragraph-Level Processing}\label{subsec:paragraph_level}
After preprocessing the OSCTI, the next phase in the \name framework is \emph{Paragraph-Level Processing}.
In this phase, \name first identifies key entities: API calls and MITRE ATT\&CK TTPs.
These entities are then used to enrich the formatted paragraphs, from which \name generates initial \sig candidates.
To perform these complex tasks, LLMs require carefully defined steps of accurate information extraction and effective data linkage.
Our experiments showed that segmenting the OSCTI text into manageable chunks (i.e., paragraphs) enhances the efficiency of the tasks involved in \sig candidate generation.
This approach is also aligned with the natural structure of writing, which organizes information into semantically distinct paragraphs, narrowing the model's focus and minimizing errors.
Additionally, we leverage parallelization by processing these paragraphs concurrently to boost processing speed significantly.

\begin{figure}[b]
  \centering
  \begin{subfigure}[h]{0.17\textwidth}
    \includegraphics[trim={0 0 1 7},clip,width=\textwidth]{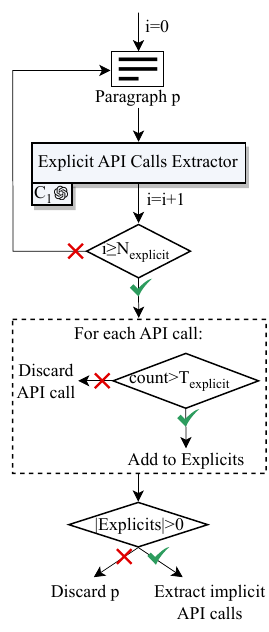}
    \caption{Explicit API calls}
    \label{fig:explicit}
  \end{subfigure}
  \hfill
  \begin{subfigure}[h]{0.17\textwidth}
    \includegraphics[trim={1 0 0 7},clip,width=\textwidth]{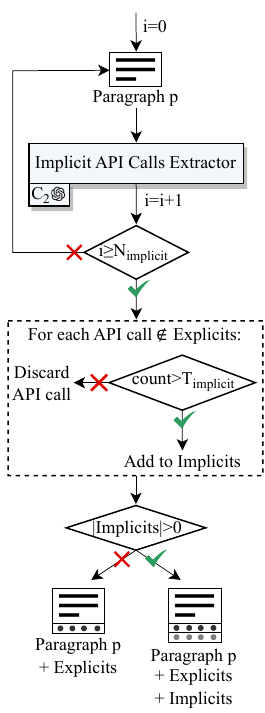}
    \caption{Implicit API calls}
    \label{fig:implicit}
  \end{subfigure}
  \caption{Threat actors' API call extraction process.}
  \label{fig:api_call_extraction}
\end{figure}

\noindent\textbf{API Call Extractor.} The \textit{Paragraph-Level Processing} phase starts with the \emph{API Call Extractor} (C in Fig.~\ref{fig:clout_overview}), which analyzes paragraphs from the OSCTI formatted text that were generated in the previous phase.
The methodology employed by this component is depicted in the flowchart presented in Fig.~\ref{fig:api_call_extraction}, which demonstrates the process of extracting API calls that are explicitly mentioned and implicitly referred to in each paragraph.
To enhance the reliability of the model's output, mitigate hallucinations (e.g., referencing events that do not exist), and prevent the omission of API calls, we incorporate a majority voting mechanism. 
This approach ensures a higher degree of accuracy and confidence in the identification and extraction of relevant API calls from the OSCTI text.

The operational flow begins with the \textit{explicit API call extractor}, where a dedicated prompt instructs the LLM to extract all explicitly mentioned API calls in the paragraph.
This operation is executed \(N_{explicit}\) times, with API calls that exceed the \(T_{explicit}\) threshold selected for subsequent analysis.
Only paragraphs containing API calls that meet the \(T_{explicit}\) are kept; the rest are discarded.

Then, paragraphs that are found to contain explicit API calls undergo more nuanced extraction by the \textit{Implicit API Call Extractor}.
In this step, we utilized the LLM to perform a deeper analysis to infer API calls suggested indirectly by the OSCTI author. 
For example, operational descriptions such as performing a \textit{sync} action on an S3 bucket should be mapped to the \textit{ListBuckets} and \textit{GetObject} API calls.
Due to the complexity of identifying these implicit API calls, this step is executed $N_{\text{implicit}}$ times, where $N_{\text{implicit}}$ is set to twice the number of $N_{\text{explicit}}$ iterations performed. 
Similar to the explicit API call extraction process, paragraphs are analyzed for implicit API calls that meet the \(T_{implicit}\) threshold.
However, paragraphs without any implicit API calls are not discarded, as they still have some value due to their explicit API call content.

\noindent\textbf{TTP Extractor.} 
This component (D in Fig.~\ref{fig:clout_overview}) analyzes the extracted API calls, mapping them to cloud-based MITRE ATT\&CK tactics, techniques, and sub-techniques. 
It utilizes a detailed prompt, which includes mapping cloud tactics to techniques and techniques to sub-techniques (in JSON format), along with illustrative examples of effective and ineffective mappings.
This integrated approach not only enhances the accuracy of TTP assignments but also safeguards against model hallucinations.
Each API call is evaluated in its specific context to assign the most precise and relevant TTPs.
While these TTPs do not directly alter the detection logic of the \sig candidates, they play a critical role in understanding the structure of the attack and classifying its various stages.
This understanding enables the subsequent \textit{Rule Generator} component to focus on detection logic (rather than on extracting additional information) and correctly dividing and classifying the rules.

\begin{figure*}
  \centering
  \begin{subfigure}[t]{0.296\textwidth}
    \includegraphics[width=\textwidth]{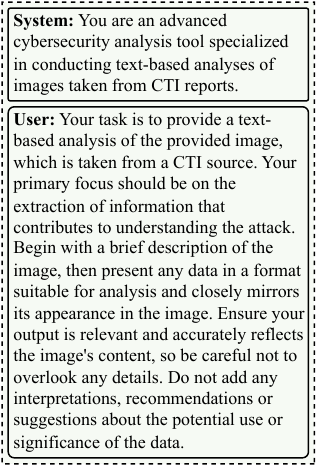}
    \caption{Image Analyzer prompt}
    \label{fig:image_analysis_prompt}
  \end{subfigure}
  \begin{subfigure}[t]{0.447\textwidth}
    \includegraphics[width=\textwidth]{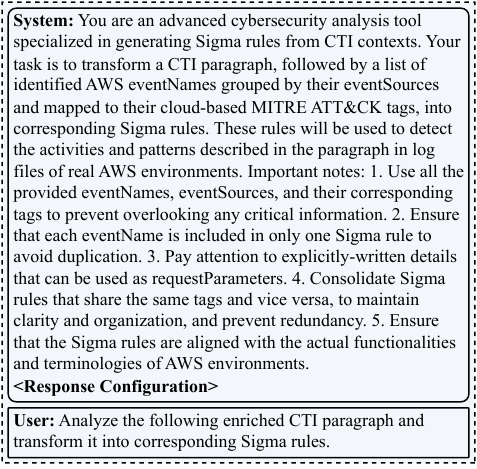}
    \caption{Rule Generator prompt}
    \label{fig:rule_generating_prompt}
  \end{subfigure}
  \begin{subfigure}[t]{0.247\textwidth}
    \includegraphics[width=\textwidth]{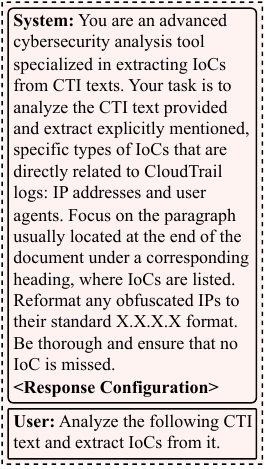}
    \caption{IoC Extractor prompt}
    \label{fig:ioc_extracting_prompt}
  \end{subfigure}
  \caption{Examples of \name prompts.}
  \label{fig:prompts}
\end{figure*}

\noindent\textbf{Rule Generator.} The last component in the \textit{Paragraph-Level Processing} phase (E in Fig.~\ref{fig:clout_overview}) is considered the core component of the \name framework.
The \textit{Rule Generator} receives a list of identified API calls along with their corresponding TTP assignments, which is bundled with the formatted paragraph text as input.
The LLM processes this enriched input using the \textit{Rule Generator} prompt, as shown in Fig.~\ref{fig:rule_generating_prompt}.
This prompt defines the LLM's role as a cybersecurity analysis tool that specializes in generating \sig rules from OSCTI text.
This approach aims to leverage extracted AWS API calls to enrich paragraphs and transform them into \sig candidates. 
This, in turn, enables the detection of similar activities or patterns in log files.
The generation prompt includes several important instructions:
\begin{itemize}[nosep,noitemsep,leftmargin=*]
    \item Each API call provided (along with its TTPs) must be included in the \sig candidates, but not more than once, to avoid the omission of important details and duplications.
    \item Paying attention to small details is extremely important as they can improve the detection specificity of the \sig candidates.
    \item \sig candidates with the same attack patterns and stages (i.e., their TTPs) should be merged and vice versa.
    \item \sig candidates must align with the specific terminology and functionality of AWS environments to ensure relevance.
   \item The output (i.e., LLM response) is required to be in a uniform and interpretable format. We used JSON format since it is a built-in feature available through the OpenAI API~\cite{openai_json_mode}.
   Note that \textit{<Response Configuration>} in Fig.~\ref{fig:rule_generating_prompt} and Fig.~\ref{fig:ioc_extracting_prompt} is a placeholder for the detailed description of the requested JSON format specified in the prompt.
\end{itemize}

\noindent\textbf{Rule Validator.} 
This component (F in Fig.~\ref{fig:clout_overview}) is applied to the \sig candidates generated by both the \textit{Rule Generator} and \textit{Rule Optimizer}.
Each candidate is examined using the following approaches to ensure uniformity and compliance with the \sig structure:
\begin{itemize}[nosep,noitemsep,leftmargin=*]
\item \textbf{Sanitation} - removes overly specific or extraneous fields that could limit the candidate's effectiveness in various contexts. 
This includes the elimination of fields such as \textit{errorcode}, \textit{errormessage}, and \textit{eventtime}, as well as any explicit resources specified in the OSCTI by their unique Amazon resource name.
By stripping out such details, this component enhances the applicability of the candidates across multiple scenarios and environments.
\item \textbf{Reformatting} - adjusts the candidate's syntax to conform to the \sig standard structure, ensuring the validity of \textit{<key: value>} pairs.
This process simplifies complex data structures mistakenly generated by the LLM, transforming nested lists and dictionaries into appropriate formats.
\item \textbf{Metadata Validating} - ensures the correctness and relevance of metadata including author name, reference URLs, and dates.
\end{itemize}
By eliminating redundant attributes and correcting flaws in the \sig candidates' structure, this component safeguards their structural integrity and consistency.

\vspace{-0.1cm}
\subsection{OSCTI-Level Processing}\label{subsec:oscti_level}
The final phase in the \name framework aggregates \sig candidates generated from individual paragraphs to enable holistic processing and enrichment.
This phase takes as input the collected \sig candidates from all previously processed paragraphs, and its output is the final, optimized set of \sig candidates which are free of redundancies and enriched with IoCs.

\noindent\textbf{Rule Optimizer.} The first component (G in Fig.~\ref{fig:clout_overview}) in the \textit{OSCTI-Level Processing} phase is designed to improve \sig candidates' detection logic.
In this component, the LLM processes the validated \sig candidates concurrently to enhance the speed and efficiency of the optimization process.
The LLM is instructed by a designated prompt and provided with examples of how to perform the optimization, ensuring that the detection criteria are both clear and aligned with their intended purpose. The optimization process includes the following tasks:

\begin{itemize}[nosep,noitemsep,leftmargin=*]
    \item \textbf{Unification} - merges \textit{selection} fields that match identical detection criteria, i.e., those sharing the same filtering logic.
    For example, consider the \sig rule in Listing~\ref{lst:sigma_candidate_example}, which detects access to a certain file from malicious IP addresses.
    Assume this \sig rule includes another \textit{selection} field with the same event source, event name, and request parameter (\textit{s3.amazonaws.com}, \textit{GetObject}, and \textit{terraform.tfstate}, respectively) but adds an additional request parameter: \textit{requestParameters.bucket: Starak}.
    When performing the unification task, the \textit{Rule Optimizer} combines these two \textit{selection} fields into a single \textit{selection} that encompasses all relevant fields: \textit{eventSource}, \textit{eventName}, \textit{requestParameters.key}, and \textit{requestParameters.bucket}.
    This unification ensures that the rules are comprehensive and free of redundancy by merging overlapping criteria while preserving their original integrity.

    \item \textbf{Separation} - Splits disjoint \textit{selection} fields that share some detection criteria but have misaligned logic.
    For example, consider the \sig rule in Listing~\ref{lst:sigma_candidate_example}.
    Assume that the initial \sig rule incorrectly included two additional unrelated fields: \textit{eventSource: iam.amazonaws.com} and \textit{eventName: PutUserPolicy} in the same existing \textit{selection} field.
    The \textit{Rule Optimizer} would recognize that these fields are unrelated to the original detection logic and would separate them into a new \textit{selection} field.
   Then, it would update the \textit{condition} field to search for either the first \textit{selection} or the newly created second \textit{selection}.
    This separation ensures that the rule remains accurate and logically consistent by distinguishing between different detection criteria.
\end{itemize}

\noindent\textbf{Set Refiner.} This component (H in Fig.~\ref{fig:clout_overview}) plays an important role in refining the overall \sig candidate set.
Essentially, it eliminates redundancies caused by paragraphs' independent processing, which often leads to the identification of duplicate API call names across different \sig candidates.
This component's logic is encapsulated in Algorithm~\ref{alg:rules_set_refining}. 
Initially, the algorithm parses through all candidates to compile a unique set of API calls. 
Subsequently, for each API call, the algorithm aggregates all corresponding candidates into a collection termed \textit{commonRules}. 
A selection prompt is then employed to identify the most suitable candidate for detecting the specified API call. 
Once a candidate is chosen, the remaining candidates, now referred to as \textit{rulesToAdjust}, undergo a process to either have the specific API call removed or, if a candidate is exclusively based on a single API call and not selected, it is discarded in its entirety.
\sig candidates containing more than one API call are subject to adjustment through \textit{RemoveAPI}, a process tailored to refine the rule by excluding the API call.
This procedure, applied to all API calls, results in a final set of \sig candidates free of redundancy, with each candidate contributing uniquely to the detection logic and covering specific types of events (with no overlapping between candidates in terms of the events covered).

\begin{algorithm}[h]
    \caption{API Call Duplication Remover.}
    \label{alg:rules_set_refining}
    \small
    \begin{flushleft}
        \textbf{Input:} A set of \sig candidates $osctiRules$\\
        \textbf{Output:} Modified $osctiRules$\\
    \end{flushleft}
    \begin{algorithmic}[1]
        \State $osctiAPIs \gets $ ExtractAPIs($osctiRules$)
        \For{each $osctiAPI \in osctiAPIs$}
            \State $commonRules \gets $ GetCommonRules($osctiRules$, $osctiAPI$)
            \State $selectedRule \gets $ SelectRule($commonRules$, $osctiAPI$)
            \State $rulesToAdjust \gets commonRules - selectedRule$
            \For{each $ruleToAdjust \in rulesToAdjust$}
                \State $ruleAPIs \gets $ ExtractAPIs($ruleToAdjust$)
                \If{$|ruleAPIs| = 1$}
                    \State $osctiRules \gets osctiRules - ruleToAdjust$
                \Else
                    \State RemoveAPIFromRule($ruleToAdjust$, $osctiAPI$)
                \EndIf
            \EndFor
        \EndFor
    \end{algorithmic}
\end{algorithm}

\noindent\textbf{IoC Extractor.} This component (I in Fig.~\ref{fig:clout_overview}) is tasked with parsing OSCTI texts to identify and extract explicit IoCs, notably IP addresses and user agents pertinent to AWS CloudTrail logs.
The input for this component is the full version of the OSCTI \textit{Downloader and Parser} component output, accompanied by a guiding prompt, as detailed in Fig.~\ref{fig:ioc_extracting_prompt}.
This prompt directs the LLM to focus on paragraphs of the OSCTI text containing IoCs, which are typically located at the end of the text (for example, in the conclusion, findings, or IoC paragraphs).
Additionally, the LLM is instructed to ensure that no IoC is overlooked and to convert obfuscated IP addresses and user agents to a standardized format.

\noindent\textbf{IoC Enhancer.} Following the extraction of IoCs by the \textit{IoC Extractor}, this component (J in Fig.~\ref{fig:clout_overview}) acts as the final step in the \textit{OSCTI-Level Processing} phase of the \name framework.
Its role is to carefully integrate the extracted IoCs into \sig candidates, thereby enhancing their threat detection capabilities without disrupting their operational logic.
For example, consider a \sig candidate that includes a single \textit{selection} field consisting of API calls and event sources.
It needs to integrate two types of IoCs: an IP address (\textit{198.51.100.1}) and a user agent (\textit{Mozilla/5.0}).
The \textit{IoC Enhancer} introduces two new \textit{selection} fields: one for the \textit{sourceIPAddress: 198.51.100.1} criteria and another for the \textit{userAgent|contains: Mozilla/5.0} criteria.
The addition of `contains' improves string matching flexibility, facilitating the inclusion of specified patterns or variations.
Following the introduction of these new IoC \textit{selection} fields, this component updates the condition logic of the \sig candidate from \textit{selection} to \textit{selection and (selection\_ip\_address or selection\_user\_agent)}.
This update logically links these \textit{selection}s, requiring the presence of the defined API calls and either IoC, thus tuning the candidate's detection capabilities with focused precision.

\section{Evaluation}\label{sec:evaluation}
In this section, we describe the creation of an annotated benchmark dataset and present the methodology and metrics used to evaluate the efficacy and accuracy of the Sigma candidates generated by \name. 
We present the results of our evaluation, which also includes an ablation study in which we analyze the impact of each of the framework's components on the overall performance.

\subsection{Dataset}
\label{subsec:evaluation_dataset}
\begin{figure}[h]
  \includegraphics[trim={0 1 0 1.5},clip,width=0.80\columnwidth]{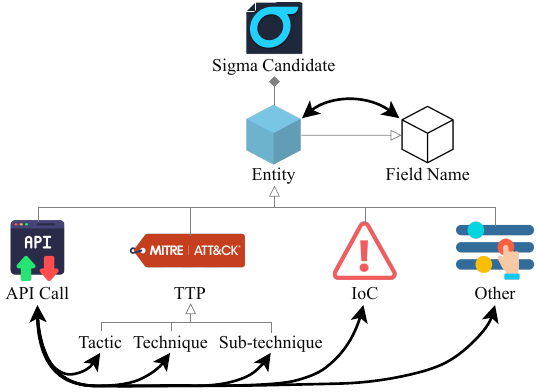}
  \caption{Evaluation setup.}
  \label{fig:evaluation_setup}
\end{figure}

We collected 12 cloud environment OSCTIs published by different vendors. 
A description of the OSCTIs, including the number of images, token size, number of API calls, and their technical complexity, is provided in Table~\ref{tab:osctis} in Appendix~\ref{app:osctis}.
To create the ground truth for the dataset, our research team (consisting of threat hunting and cloud security experts) conducted a thorough analysis of each OSCTI's content to (1) identify and extract the entities described in the OSCTI, and (2) identify the relevant inter-entity relationships essential for the creation of coherent and meaningful \sig candidates (the list of extracted entities and inter-entity relationships is provided in Table~\ref{tab:entity_relationships} and illustrated in Fig.~\ref{fig:evaluation_setup}).

\begin{table}[h]
    \centering
    \begin{adjustbox}{width=0.65\columnwidth,center}
        \begin{tabular}{|c|c|}
            \hline
            \textbf{Entity} & \textbf{Relationship} \\
            \hline
            API Call & Detection Entity $\leftrightarrow$ Sigma Field \\
            Tactic & API Call $\leftrightarrow$ Tactic \\
            Technique & API Call $\leftrightarrow$ Technique \\
            Sub-technique & API Call $\leftrightarrow$ Sub-technique \\
            IoC & API Call $\leftrightarrow$ IoC \\
            Other & API Call $\leftrightarrow$ Other \\
            \hline
        \end{tabular}
    \end{adjustbox}
    \caption{Entity types and relationships.}
    \label{tab:entity_relationships}
\end{table}
\vspace{-1cm}

\vspace{-0.5cm}
\subsection{Evaluation Metrics}
\label{subsec:evaluation_metrics}
We evaluated our framework’s performance using common entity and relationship extraction metrics, as done in prior studies~\cite{gao2021enabling,satyapanich2020casie,satvat2021extractor,sarhan2021open,park2022full,fujii2022cyner,alam2023looking,purba2023extracting,siracusano2023time,liu2023constructing,ahmed2024cyberentrel,fengrui2024few}, and the results were validated against the ground truth defined by our research team. 
Additionally, we defined a set of criteria specifically designed to test each \sig candidate’s functionality in the operational context of OSCTI. 

The metrics used to assess \name's performance in extracting and identifying the entities and inter-entity relationships in the OSCTI are the precision (P), recall (R), and F1 score (F1) weighted by the total number of entities/relationships of each type, denoted as `\#' (since each OSCTI has a different number of entities/relationships).
Calculating these metrics separately for each entity and relationship type allows us to pinpoint areas of strength and opportunities for improvement.

To evaluate the functionality, logical validity, and relevance of the \sig candidates generated by \name, we defined the following criteria.
These metrics were calculated by our research team for each \sig candidate generated:
\begin{itemize}[nosep,leftmargin=*]
    \item \textbf{Syntax Correctness} - Assesses whether the generated \sig candidates are syntactically correct and properly formatted, ensuring that a given rule is operational in a SIEM system. 
    This is achieved by compiling the rules using \sig CLI~\cite{sigma_cli}, a scripting tool that allows for the compilation and conversion of rules into query languages (e.g., Splunk).
    \item \textbf{Condition Field Accuracy} - Focuses on the correctness of the \textit{condition} fields of the \sig candidates, which specify the relationship between the various \textit{selection} fields (each of which includes a different detection artifact).
    \item \textbf{Criticality Accuracy} - Measures the accuracy of the \textit{level} field of each \sig candidate, which represents the level of importance and urgency of the rule.
    \item \textbf{Descriptive Metadata Alignment} - Evaluates whether the \textit{title}, \textit{description}, and \textit{falsepositives} fields of the \sig candidates accurately reflect their intended purpose and context.
\end{itemize}

\vspace{-0.5cm}
\subsection{Results}\label{subsec:results}
\textbf{Entity and Relationship Extraction.}
The results of our entity and relationship extraction evaluation are presented in Tables~\ref{tab:entity_extraction_results} and ~\ref{tab:relationship_extraction_results}, respectively.
We consider API calls and IoCs to be the most important entities for generating practical and relevant \sig candidates.
For these two entity types, \name achieved a weighted average precision of 92\% with a recall of 98\% for the API calls and a precision of 99\% with a recall of 98\% for the IoCs.
In the `Other' entity category (which includes entities such as request parameters and IP address, the weighted precision and recall values were 88\% and 82\%, respectively.
\name's overall performance in the task of entity extraction is a weighted precision value of 80\% and recall value of 83\%.
The reason for the reduction in the overall performance lies in the difficulty of identifying and extracting MITRE ATT\&CK TTPs~\cite{pyramid_of_pain}.
While for OSCTIs 1,6,8, and 9, \name was able to achieve high precision and recall values (over 80\%), it was less successful in extracting the correct tactic for the other OSCTIs.
It is important to note that while identifying the TTPs for the candidates generated can assist security analysts in researching and analyzing potential security incidents, it does not impact the \sig candidates' detection capabilities.

The relationship identification results, which represent \name's ability to interrelate non-TTP entities, achieved a weighted precision value of 88\% and a weighted recall value of 89\%.
When including TTP-related relationships, the overall weighted precision and recall are 75\% and 78\%, respectively, reflecting the challenges in accurately identifying these complex relationships.
\name's ability to accurately identify the \textit{Detection Entity $\leftrightarrow$ \sig Field} relationship, which ensures that non-TTP entities are correctly labeled with appropriate keys in the \sig candidates, achieved precision value of 87\% and recall value of 90\%.
The remaining relationships emphasize the connections between the primary detection entity, API calls, and other entity types.

In summary, \name demonstrates strong performance in extracting and identifying key entities and their relationships within OSCTI.
While the framework was shown to excel in handling API calls, IoCs, and request parameters, achieving high precision and recall for this, it faces challenges with MITRE ATT\&CK TTPs, which impacts the overall performance but does not affect the detection capabilities of the \sig candidates generated.

\textbf{Sigma Candidate Evaluation.}
The results of our \sig candidate evaluation are presented in Table~\ref{tab:sigma_candidate_evaluation_results}.
Promising results were obtained in this evaluation which demonstrated \name's ability to produce syntactically correct (executable) \sig rules. 
Our evaluation shows that 99.18\% of the candidates generated (121 of the 122 \sig candidates generated) were successfully compiled and converted into the Splunk query language. 
On the proposed metric of condition field accuracy, \name achieved a perfect score of 100\%, correctly specifying the logical connection between the various selection fields for all \sig candidates.
However, on the criticality accuracy metric, the \sig candidates obtained a weighted average of 75.41\%, indicating variability across different OSCTIs and highlighting an area for improvement.
Finally, the value obtained for the descriptive alignment metadata metric was high, with a weighted average of 95.90\%, resulting in an accurate description of each candidate.

\begin{table}[h]
    \centering
    \begin{adjustbox}{width=0.999\columnwidth,center}
        \begin{tabular}{|c|c|c|c|c|c|}
            \hline
            \multirow{3}{*}{\textbf{OSCTI ID}} & \multirow{3}{*}{\textbf{\#Rules}} & \multirow{3}{*}{\textbf{Executability}} & \textbf{Condition} & \multirow{2}{*}{\textbf{Criticality}} & \textbf{Descriptive} \\
                                               &                                   &                                         & \textbf{Field}     & \multirow{2}{*}{\textbf{Accuracy}}    & \textbf{Metadata}    \\
                                               &                                   &                                         & \textbf{Accuracy}  &                                       & \textbf{Alignment}   \\
            \hline
            \textbf{1}                         & 7                                 & 7 (100\%)                               & 7 (100\%)          & 6 (85.71\%)                           & 7 (100\%)            \\
            \hline 
            \textbf{2}                         & 6                                 & 6 (100\%)                               & 6 (100\%)          & 4 (66.67\%)                           & 6 (100\%)            \\
            \hline
            \textbf{3}                         & 13                                & 13 (100\%)                              & 13 (100\%)         & 12 (92.31\%)                          & 12 (92.31\%)         \\
            \hline
            \textbf{4}                         & 12                                & 12 (100\%)                              & 12 (100\%)         & 10 (83.33\%)                          & 12 (100\%)           \\
            \hline
            \textbf{5}                         & 9                                 & 9 (100\%)                               & 9 (100\%)          & 7 (77.78\%)                           & 9 (100\%)            \\
            \hline
            \textbf{6}                         & 4                                 & 4 (100\%)                               & 4 (100\%)          & 2 (50.00\%)                           & 3 (75.00\%)          \\
            \hline
            \textbf{7}                         & 11                                & 11 (100\%)                              & 11 (100\%)         & 9 (81.82\%)                           & 11 (100\%)           \\
            \hline
            \textbf{8}                         & 13                                & 13 (100\%)                              & 13 (100\%)         & 9 (69.23\%)                           & 12 (92.31\%)         \\
            \hline
            \textbf{9}                         & 14                                & 14 (100\%)                              & 14 (100\%)         & 10 (71.43\%)                          & 13 (92.86\%)         \\
            \hline
            \textbf{10}                        & 9                                 & 9 (100\%)                               & 9 (100\%)          & 7 (77.78\%)                           & 9 (100\%)            \\
            \hline
            \textbf{11}                        & 11                                & 10 (90.91\%)                            & 11 (100\%)         & 8 (72.73\%)                           & 10 (90.91\%)         \\
            \hline
            \textbf{12}                        & 13                                & 13 (100\%)                              & 13 (100\%)         & 8 (61.54\%)                           & 13 (100\%)           \\
            \hline
            \hline
            \textbf{Weighted Avg.}             & 10.17                             & 99.18\%                                 & 100.00\%           & 75.41\%                               & 95.90\%              \\
            \hline
        \end{tabular}
    \end{adjustbox}
    \caption{Sigma candidate evaluation results.}
    \label{tab:sigma_candidate_evaluation_results}
\end{table}

\vspace{-0.7cm}
\textbf{Ablation Study Results.}
We conducted an ablation study to better understand the impact of \name's components on its performance. 
To do this, we created three variations of \name by systematically removing key components and evaluating the performance of each variant. 
Table~\ref{tab:ablation_study_configurations} in Appendix~\ref{app:ablation} summarizes the different configurations used in the ablation study. 
The \noimagesmodel variation evaluates the impact of the \textit{Image Analyzer} component (B in Fig.~\ref{fig:clout_overview}); the \noapimodel variation is designed to evaluate the impact of the \textit{API Call Extractor} and \textit{TTP Extractor} components (C and D in Fig.~\ref{fig:clout_overview}, respectively); and the \nooptimizationmodel variation aims to evaluate the \textit{Rule Optimizer} component (G in Fig.~\ref{fig:clout_overview}).
Table~\ref{tab:ablation_study_results} in Appendix~\ref{app:ablation} presents the results for each of the variations in the previously evaluated entity and relationship identification tasks.

The results obtained when using the \noimagesmodel variation show a 7.66\% drop in weighted average precision and, importantly, a 15.33\% drop in the recall for the entity extraction task.
Additionally, the weighted average precision and recall for relationship extraction were reduced by 10.66\% and 23.67\%, respectively.
This significant reduction in accuracy, especially in extraction coverage, highlights the importance of the \textit{Image Analyzer} component in extracting information from images that may not be available elsewhere.

The \noapimodel variation, with the \textit{API Call Extractor} and \textit{TTP Extractor} components removed, resulted in significantly worse performance compared to the other variations.
For the task of entity extraction, we observed a 30.83\% drop in the average precision and a 22.66\% drop in the average recall.
Performance on the relationship extraction metrics was even more affected, with a 37.83\% reduction in the average precision and a 29.5\% reduction in the average recall.
These findings highlight the importance of separating entity extraction from rule generation so that the model can focus on one task at a time.
Specifically, the \textit{API Call Extractor} and \textit{TTP Extractor} components proved essential to \name's overall performance.
In contrast, less dramatic differences in the performance were seen with the \nooptimizationmodel variation, which assesses the impact of omitting the \textit{Rule Optimizer} component. 
Average entity extraction precision and recall dropped by 5.16\% and 6.66\%, respectively.
In the relationship extraction task, there was a 6.33\% reduction in average precision and a 7.5\% decrease in average recall.
Although these declines are not as great as those seen in the previous variation, they indicate that the \textit{Rule Optimizer} plays a meaningful role in \name's performance.

To summarize, the ablation study highlights the essential roles of the \textit{Image Analyzer}, \textit{API Call Extractor}, and \textit{TTP Extractor} components in maintaining high precision and recall in both entity and relationship extraction tasks.
The \textit{Rule Optimizer} also plays a valuable role, though its impact is less pronounced compared to the other components.

\begin{table*}
    \centering
    \begin{adjustbox}{width=0.999\textwidth,center}
        \begin{tabular}{|*{25}{c|}}
            \hline
            \textbf{OSCTI} & \multicolumn{4}{c|}{\textbf{API Call}} & \multicolumn{4}{c|}{\textbf{Tactic}} & \multicolumn{4}{c|}{\textbf{Technique}} & \multicolumn{4}{c|}{\textbf{Sub-technique}} & \multicolumn{4}{c|}{\textbf{IoC}} & \multicolumn{4}{c|}{\textbf{Other}} \\
            \hhline{~*{24}{-}}
            \textbf{ID} & \textbf{\#} & \textbf{P} & \textbf{R} & \textbf{F1} & \textbf{\#} & \textbf{P} & \textbf{R} & \textbf{F1} & \# & \textbf{P} & \textbf{R} & \textbf{F1} & \textbf{\#} & \textbf{P} & \textbf{R} & \textbf{F1} & \textbf{\#} & \textbf{P} & \textbf{R} & \textbf{F1} & \# & \textbf{P} & \textbf{R} & \textbf{F1} \\
            \hline
            \textbf{1} & 8 & \gradient{1.00} & \gradient{1.00} & \gradient{1.00} & 5 & \gradient{1.00} & \gradient{1.00} & \gradient{1.00} & 5 & \gradient{1.00} & \gradient{1.00} & \gradient{1.00} & 4 & \gradient{0.80} & \gradient{1.00} & \gradient{0.89} & 2 & \gradient{1.00} & \gradient{1.00} & \gradient{1.00} & 13 & \gradient{0.85} & \gradient{0.85} & \gradient{0.85} \\
            \hline
            \textbf{2} & 6 & \gradient{1.00} & \gradient{1.00} & \gradient{1.00} & 4 & \gradient{0.75} & \gradient{0.75} & \gradient{0.75} & 4 & \gradient{0.75} & \gradient{0.75} & \gradient{0.75} & 2 & \gradient{0.50} & \gradient{0.50} & \gradient{0.50} & 3 & \gradient{1.00} & \gradient{1.00} & \gradient{1.00} & 3 & \gradient{1.00} & \gradient{0.67} & \gradient{0.80} \\
            \hline
            \textbf{3} & 12 & \gradient{0.71} & \gradient{1.00} & \gradient{0.83} & 4 & \gradient{0.50} & \gradient{0.75} & \gradient{0.60} & 8 & \gradient{0.56} & \gradient{0.63} & \gradient{0.59} & 5 & \gradient{0.67} & \gradient{0.40} & \gradient{0.50} & 3 & \gradient{1.00} & \gradient{0.67} & \gradient{0.80} & 24 & \gradient{0.86} & \gradient{1.00} & \gradient{0.92} \\
            \hline
            \textbf{4} & 49 & \gradient{0.98} & \gradient{1.00} & \gradient{0.99} & 5 & \gradient{0.75} & \gradient{0.60} & \gradient{0.67} & 8 & \gradient{0.60} & \gradient{0.75} & \gradient{0.67} & 3 & \gradient{0.40} & \gradient{0.67} & \gradient{0.50} & 1 & \gradient{1.00} & \gradient{1.00} & \gradient{1.00} & 6 & \gradient{0.86} & \gradient{1.00} & \gradient{0.92} \\
            \hline
            \textbf{5} & 9 & \gradient{1.00} & \gradient{1.00} & \gradient{1.00} & 4 & \gradient{0.75} & \gradient{0.75} & \gradient{0.75} & 4 & \gradient{0.67} & \gradient{1.00} & \gradient{0.80} & 5 & \gradient{1.00} & \gradient{1.00} & \gradient{1.00} & 3 & \gradient{1.00} & \gradient{1.00} & \gradient{1.00} & 4 & \gradient{1.00} & \gradient{1.00} & \gradient{1.00} \\
            \hline
            \textbf{6} & 4 & \gradient{1.00} & \gradient{1.00} & \gradient{1.00} & 2 & \gradient{1.00} & \gradient{1.00} & \gradient{1.00} & 2 & \gradient{0.67} & \gradient{1.00} & \gradient{0.80} & 1 & \gradient{0.50} & \gradient{1.00} & \gradient{0.67} & 2 & \gradient{1.00} & \gradient{1.00} & \gradient{1.00} & 3 & \gradient{0.75} & \gradient{1.00} & \gradient{0.86} \\
            \hline
            \textbf{7} & 12 & \gradient{0.71} & \gradient{1.00} & \gradient{0.83} & 4 & \gradient{0.60} & \gradient{0.75} & \gradient{0.67} & 5 & \gradient{0.38} & \gradient{0.60} & \gradient{0.46} & 6 & \gradient{0.25} & \gradient{0.33} & \gradient{0.29} & 1 & \gradient{1.00} & \gradient{1.00} & \gradient{1.00} & 4 & \gradient{0.57} & \gradient{1.00} & \gradient{0.73} \\
            \hline
            \textbf{8} & 14 & \gradient{0.93} & \gradient{0.93} & \gradient{0.93} & 0 & \gradient{0.80} & \gradient{0.80} & \gradient{0.80} & 2 & \gradient{0.71} & \gradient{0.71} & \gradient{0.71} & 0 & \gradient{0.50} & \gradient{1.00} & \gradient{0.67} & 67 & \gradient{0.99} & \gradient{0.99} & \gradient{0.99} & 10 & \gradient{0.82} & \gradient{0.90} & \gradient{0.86} \\
            \hline
            \textbf{9} & 20 & \gradient{1.00} & \gradient{1.00} & \gradient{1.00} & 1 & \gradient{0.83} & \gradient{1.00} & \gradient{0.91} & 4 & \gradient{0.78} & \gradient{0.78} & \gradient{0.78} & 6 & \gradient{0.83} & \gradient{0.83} & \gradient{0.83} & 4 & \gradient{1.00} & \gradient{1.00} & \gradient{1.00} & 6 & \gradient{0.83} & \gradient{0.83} & \gradient{0.83} \\
            \hline
            \textbf{10} & 7 & \gradient{0.58} & \gradient{1.00} & \gradient{0.74} & 3 & \gradient{0.75} & \gradient{1.00} & \gradient{0.86} & 4 & \gradient{0.43} & \gradient{0.75} & \gradient{0.55} & 3 & \gradient{0.50} & \gradient{0.67} & \gradient{0.57} & 0 & \gradient{1.00} & \gradient{1.00} & \gradient{1.00} & 5 & \gradient{1.00} & \gradient{0.40} & \gradient{0.57} \\
            \hline
            \textbf{11} & 11 & \gradient{0.91} & \gradient{0.91} & \gradient{0.91} & 5 & \gradient{0.75} & \gradient{0.60} & \gradient{0.67} & 7 & \gradient{0.57} & \gradient{0.57} & \gradient{0.57} & 4 & \gradient{0.50} & \gradient{0.75} & \gradient{0.60} & 10 & \gradient{1.00} & \gradient{1.00} & \gradient{1.00} & 22 & \gradient{0.92} & \gradient{0.50} & \gradient{0.65} \\
            \hline
            \textbf{12} & 15 & \gradient{0.93} & \gradient{0.87} & \gradient{0.90} & 5 & \gradient{0.50} & \gradient{0.60} & \gradient{0.55} & 6 & \gradient{0.56} & \gradient{0.83} & \gradient{0.67} & 6 & \gradient{0.83} & \gradient{0.83} & \gradient{0.83} & 7 & \gradient{1.00} & \gradient{1.00} & \gradient{1.00} & 16 & \gradient{0.93} & \gradient{0.88} & \gradient{0.90} \\
            \hline
            \hline
            \textbf{Weighted} &  & \multirow{2}{*}{\textbf{0.92}} & \multirow{2}{*}{\textbf{0.98}} & \multirow{2}{*}{\textbf{0.94}} &  & \multirow{2}{*}{\textbf{0.73}} & \multirow{2}{*}{\textbf{0.76}} & \multirow{2}{*}{\textbf{0.74}} &  & \multirow{2}{*}{\textbf{0.62}} & \multirow{2}{*}{\textbf{0.75}} & \multirow{2}{*}{\textbf{0.68}} &  & \multirow{2}{*}{\textbf{0.65}} & \multirow{2}{*}{\textbf{0.71}} & \multirow{2}{*}{\textbf{0.67}} &  & \multirow{2}{*}{\textbf{0.99}} & \multirow{2}{*}{\textbf{0.98}} & \multirow{2}{*}{\textbf{0.98}} &  & \multirow{2}{*}{\textbf{0.88}} & \multirow{2}{*}{\textbf{0.82}} & \multirow{2}{*}{\textbf{0.82}} \\
            \textbf{Avg.} &  &  &  &  &  &  &  &  &  &  &  &  &  &  &  &  &  &  &  &  &  &  &  &  \\
            \hline
        \end{tabular}
    \end{adjustbox}
    \caption{Entity extraction results.}
    \label{tab:entity_extraction_results}
\end{table*}
\begin{table*}
    \vspace{-0.7cm}
    \centering
    \begin{adjustbox}{width=0.999\textwidth,center}
        \begin{tabular}{|*{25}{c|}}
            \hline
            \textbf{OSCTI} & \multicolumn{4}{c|}{\textbf{Detection Entity $\leftrightarrow$ Sigma Field}} & \multicolumn{4}{c|}{\textbf{API Call $\leftrightarrow$ Tactic}} & \multicolumn{4}{c|}{\textbf{API Call $\leftrightarrow$ Technique}} & \multicolumn{4}{c|}{\textbf{API Call $\leftrightarrow$ Sub-technique}} & \multicolumn{4}{c|}{\textbf{API Call $\leftrightarrow$ IoC}} & \multicolumn{4}{c|}{\textbf{API Call $\leftrightarrow$ Other}} \\
            \hhline{~*{24}{-}}
            \textbf{ID} & \textbf{\#} & \textbf{P} & \textbf{R} & \textbf{F1} & \textbf{\#} & \textbf{P} & \textbf{R} & \textbf{F1} & \# & \textbf{P} & \textbf{R} & \textbf{F1} & \textbf{\#} & \textbf{P} & \textbf{R} & \textbf{F1} & \textbf{\#} & \textbf{P} & \textbf{R} & \textbf{F1} & \# & \textbf{P} & \textbf{R} & \textbf{F1} \\
            \hline
            \textbf{1} & 24 & \gradient{0.83} & \gradient{0.83} & \gradient{0.83} & 8 & \gradient{0.88} & \gradient{0.88} & \gradient{0.88} & 8 & \gradient{0.88} & \gradient{0.88} & \gradient{0.88} & 6 & \gradient{0.83} & \gradient{0.83} & \gradient{0.83} & 16 & \gradient{1.00} & \gradient{1.00} & \gradient{1.00} & 18 & \gradient{0.85} & \gradient{0.94} & \gradient{0.89} \\
            \hline
            \textbf{2} & 11 & \gradient{0.82} & \gradient{0.82} & \gradient{0.82} & 6 & \gradient{0.83} & \gradient{0.83} & \gradient{0.83} & 6 & \gradient{0.83} & \gradient{0.83} & \gradient{0.83} & 3 & \gradient{0.50} & \gradient{0.67} & \gradient{0.57} & 18 & \gradient{1.00} & \gradient{1.00} & \gradient{1.00} & 10 & \gradient{1.00} & \gradient{0.70} & \gradient{0.82} \\
            \hline
            \textbf{3} & 39 & \gradient{0.81} & \gradient{0.97} & \gradient{0.88} & 12 & \gradient{0.59} & \gradient{0.83} & \gradient{0.69} & 12 & \gradient{0.40} & \gradient{0.67} & \gradient{0.50} & 5 & \gradient{0.50} & \gradient{0.40} & \gradient{0.44} & 21 & \gradient{0.59} & \gradient{0.95} & \gradient{0.73} & 32 & \gradient{0.82} & \gradient{0.97} & \gradient{0.89} \\
            \hline
            \textbf{4} & 55 & \gradient{0.95} & \gradient{1.00} & \gradient{0.97} & 49 & \gradient{0.86} & \gradient{0.88} & \gradient{0.87} & 49 & \gradient{0.53} & \gradient{0.63} & \gradient{0.57} & 5 & \gradient{0.33} & \gradient{0.40} & \gradient{0.36} & 49 & \gradient{0.98} & \gradient{1.00} & \gradient{0.99} & 49 & \gradient{0.88} & \gradient{0.90} & \gradient{0.89} \\
            \hline
            \textbf{5} & 15 & \gradient{0.94} & \gradient{1.00} & \gradient{0.97} & 9 & \gradient{0.78} & \gradient{0.78} & \gradient{0.78} & 9 & \gradient{0.67} & \gradient{0.67} & \gradient{0.67} & 8 & \gradient{0.71} & \gradient{0.63} & \gradient{0.67} & 27 & \gradient{1.00} & \gradient{1.00} & \gradient{1.00} & 10 & \gradient{1.00} & \gradient{1.00} & \gradient{1.00} \\
            \hline
            \textbf{6} & 9 & \gradient{0.70} & \gradient{0.78} & \gradient{0.74} & 4 & \gradient{1.00} & \gradient{1.00} & \gradient{1.00} & 4 & \gradient{0.67} & \gradient{0.50} & \gradient{0.57} & 3 & \gradient{0.67} & \gradient{0.67} & \gradient{0.67} & 8 & \gradient{1.00} & \gradient{1.00} & \gradient{1.00} & 4 & \gradient{0.75} & \gradient{0.75} & \gradient{0.75} \\
            \hline
            \textbf{7} & 17 & \gradient{0.68} & \gradient{1.00} & \gradient{0.81} & 13 & \gradient{0.59} & \gradient{0.77} & \gradient{0.67} & 12 & \gradient{0.45} & \gradient{0.83} & \gradient{0.59} & 10 & \gradient{0.14} & \gradient{0.30} & \gradient{0.19} & 12 & \gradient{0.71} & \gradient{1.00} & \gradient{0.83} & 12 & \gradient{0.65} & \gradient{0.92} & \gradient{0.76} \\
            \hline
            \textbf{8} & 91 & \gradient{0.97} & \gradient{0.98} & \gradient{0.97} & 14 & \gradient{0.79} & \gradient{0.79} & \gradient{0.79} & 12 & \gradient{0.54} & \gradient{0.58} & \gradient{0.56} & 5 & \gradient{0.57} & \gradient{0.80} & \gradient{0.67} & 952 & \gradient{0.91} & \gradient{0.90} & \gradient{0.91} & 20 & \gradient{0.85} & \gradient{0.85} & \gradient{0.85} \\
            \hline
            \textbf{9} & 31 & \gradient{0.93} & \gradient{0.90} & \gradient{0.92} & 20 & \gradient{0.90} & \gradient{0.90} & \gradient{0.90} & 20 & \gradient{0.65} & \gradient{0.65} & \gradient{0.65} & 12 & \gradient{0.73} & \gradient{0.67} & \gradient{0.70} & 80 & \gradient{1.00} & \gradient{1.00} & \gradient{1.00} & 20 & \gradient{0.86} & \gradient{0.90} & \gradient{0.88} \\
            \hline
            \textbf{10} & 12 & \gradient{0.64} & \gradient{0.75} & \gradient{0.69} & 7 & \gradient{0.42} & \gradient{0.71} & \gradient{0.53} & 7 & \gradient{0.50} & \gradient{0.86} & \gradient{0.63} & 5 & \gradient{0.71} & \gradient{1.00} & \gradient{0.83} & 0 & \gradient{1.00} & \gradient{1.00} & \gradient{1.00} & 10 & \gradient{0.58} & \gradient{0.70} & \gradient{0.64} \\
            \hline
            \textbf{11} & 45 & \gradient{0.74} & \gradient{0.62} & \gradient{0.67} & 12 & \gradient{0.67} & \gradient{0.50} & \gradient{0.57} & 12 & \gradient{0.67} & \gradient{0.50} & \gradient{0.57} & 6 & \gradient{0.43} & \gradient{0.50} & \gradient{0.46} & 110 & \gradient{0.91} & \gradient{0.91} & \gradient{0.91} & 29 & \gradient{0.76} & \gradient{0.55} & \gradient{0.64} \\
            \hline
            \textbf{12} & 38 & \gradient{0.94} & \gradient{0.87} & \gradient{0.90} & 15 & \gradient{0.64} & \gradient{0.60} & \gradient{0.62} & 15 & \gradient{0.40} & \gradient{0.40} & \gradient{0.40} & 11 & \gradient{0.63} & \gradient{0.45} & \gradient{0.53} & 105 & \gradient{0.93} & \gradient{0.87} & \gradient{0.90} & 25 & \gradient{0.91} & \gradient{0.84} & \gradient{0.88} \\
            \hline
            \hline
            \textbf{Weighted} &  & \multirow{2}{*}{\textbf{0.87}} & \multirow{2}{*}{\textbf{0.90}} & \multirow{2}{*}{\textbf{0.88}} &  & \multirow{2}{*}{\textbf{0.77}} & \multirow{2}{*}{\textbf{0.80}} & \multirow{2}{*}{\textbf{0.78}} &  & \multirow{2}{*}{\textbf{0.56}} & \multirow{2}{*}{\textbf{0.64}} & \multirow{2}{*}{\textbf{0.59}} &  & \multirow{2}{*}{\textbf{0.56}} & \multirow{2}{*}{\textbf{0.58}} & \multirow{2}{*}{\textbf{0.56}} &  & \multirow{2}{*}{\textbf{0.92}} & \multirow{2}{*}{\textbf{0.91}} & \multirow{2}{*}{\textbf{0.92}} &  & \multirow{2}{*}{\textbf{0.84}} & \multirow{2}{*}{\textbf{0.85}} & \multirow{2}{*}{\textbf{0.84}} \\
            \textbf{Avg.} &  &  &  &  &  &  &  &  &  &  &  &  &  &  &  &  &  &  &  &  &  &  &  &  \\
            \hline
        \end{tabular}
    \end{adjustbox}
    \caption{Relationship extraction results.}
    \label{tab:relationship_extraction_results}
\end{table*}

\section{\label{sec:discussion}Discussion}
In this section, we present the insights gained during \name's development and evaluation and our recommendations regarding the use of LLMs to automate complex NLP tasks in the field of cybersecurity.
Our experiments highlighted the effectiveness of the different techniques used in our research, which we applied across \name's pipeline.
These details, outlining the purpose and specific settings for each component, are summarized in Table~\ref{tab:component_configurations} in Appendix~\ref{app:component_configurations} and described below:

\noindent\textbf{Majority Rule in Entity Extraction Using LLMs.}
In the \textit{API Call Extractor} component we employ a majority voting mechanism to address LLMs' inconsistency. 
While the same extraction request to an LLM generally yields similar results, variations can occur due to their generative nature. 
To increase the confidence and accuracy of extracted API calls, we use majority voting where only extracted API calls meeting a specified threshold are considered. 
Since each LLM request incurs cost and runtime, we conducted experiments to determine the optimal tradeoff between the number of runs, the threshold size, and the accuracy of the results.
Our experiments showed that this voting mechanism was effective in ambiguous cases and decreased model hallucinations.

\noindent\textbf{Structured Response Format.}
For each LLM request, we use the JSON output format via the \textit{<Response Configuration>} setting. 
This structured format enables automatic validation and processing of responses. 
It also allows direct access to values in the generated text without the need for additional postprocessing of the response.

\noindent\textbf{LLM Temperature Settings.}
The temperature setting of an LLM influences the creativity and randomness of its outputs, and its values range between zero and two~\cite{openai_temperature}. 
By adjusting the temperature for different tasks, we can improve the results. 
For example, in the \textit{API Call Extractor} component, where extracting the information accurately is crucial, we use a low temperature of zero to ensure more accurate responses. 
In contrast, for the \textit{Rule Generator} component, we set the temperature to 0.7 to allow the model to generate conditions for \sig rules, which require some `creativity.'

\noindent\textbf{Leveraging the Few-shot Learning Technique.}
By providing instructions and examples for input and output, we can significantly improve the model's performance~\cite{brown2020language,ouyang2022training}. 
Since we divided the task of converting OSCTI text into smaller, well-defined tasks, we were able to provide specific instructions to the model for each of the tasks. 
Using few-shot learning, we provided a small number of examples along with the instructions, which further enhanced the model's ability to generate accurate and relevant outputs.

\noindent\textbf{Parallel LLM Requests.}
We leveraged independent LLM prompts to perform parallel execution, resulting in improved speed and efficiency. 
We identified two key scenarios where parallel requests were particularly beneficial. 
First, in the preprocessing phase, we sent all images to be translated into text simultaneously, significantly accelerating this step. 
Second, in the paragraph-level processing phase, we applied parallel processing to each paragraph, which greatly reduced the overall processing time, a threefold decrease in time was observed. 
This approach not only reduces the runtime but also improves scalability for larger datasets, allowing for more efficient handling of extensive text corpora.

\noindent\textbf{Limitations.}
In our work, we identified several factors that could affect the performance of our method.
While some of these challenges are inherently due to existing technological and methodological constraints, others present opportunities for future work. 
First, we used a commercial LLM model (OpenAI's \model) due to its high performance and user-centric benchmark results~\cite{wang2024user, gpt_4o_analysis}. 
This introduces a cost factor that needs to be considered.
Second, the method assumes that OSCTIs are divided into paragraphs. 
This limits the proposed method to use OSCTIs that are not structured in this way.
Finally, while we used pretrained LLMs, fine-tuning open-source models, may have an advantage in performing specific tasks correctly.

\vspace{0.3cm}
\section{Conclusions and Future Work}
In this paper, we presented \name, an end-to-end framework that analyzes textual and visual OSCTI using a pretrained LLM model when provided a URL. 
Our framework offers significant flexibility by allowing easy updates to newer and improved models without the need for fine-tuning, and it demonstrates scalability by running independently across multiple OSCTI images and paragraphs. 
By using the \sig format, \name's output can be seamlessly integrated into existing SIEM systems.
Future work can focus on extending \name to on-premise environments, increasing its applicability in diverse organizational settings and environments.
Additionally, we plan to enhance our framework by equipping it with playbook automation capabilities, which will improve its ability to mitigate detected threats and provide more robust support for threat hunters.

\clearpage
\bibliographystyle{ACM-Reference-Format}
\bibliography{bibfile}

\appendix
\section{OSCTI Sources Used in our Research}\label{app:osctis}

In Table~\ref{tab:osctis}, we present the list of OSCTI sources used in our development and evaluation of \name.
For each source, we present the number of images included, the number of tokens (which serve as the input to the LLM), the number of API calls, and our rating of the technical complexity of the OSCTI.

\begin{table*}[h]
    \centering
    \begin{adjustbox}{width=0.7\textwidth,center}
        \begin{tabular}{|c|c|c|c|c|c|c|}
            \hline
            \textbf{OSCTI} & \multirow{2}{*}{\textbf{OSCTI Name}} & \multirow{2}{*}{\textbf{\#Images}} & \multicolumn{2}{c|}{\textbf{\#Tokens}} & {\textbf{\#API}} & {\textbf{Technical}} \\
            \cline{4-5}
            \textbf{ID} &  &  & \textbf{No Images} & \textbf{Images} & {\textbf{Calls}} & {\textbf{Complexity}} \\
            \hline
            1 & \begin{tabular}[c]{@{}c@{}}Anatomy of an Attack: \\ Exposed keys to Crypto Mining\end{tabular} & 1 & 959 & 1,429 & 7 & High \\
            \hline
            2 & \begin{tabular}[c]{@{}c@{}}Behind the scenes in the Expel SOC: \\ Alert-to-fix in AWS\end{tabular} & 5 & 2,822 & 5,275 & 7 & Medium \\
            \hline
            3 & \begin{tabular}[c]{@{}c@{}}CloudKeys in the Air: \\ Tracking Malicious Operations of Exposed IAM Keys\end{tabular} & 10 & 4,857 & 10,393 & 10 & Low \\
            \hline
            4 & \begin{tabular}[c]{@{}c@{}}Compromised Cloud Compute Credentials \\ Case Studies From the Wild\end{tabular} & 5 & 3,108 & 4,660 & 49 & Low \\
            \hline
            5 & \begin{tabular}[c]{@{}c@{}}Incident report: From CLI to console, \\ chasing an attacker in AWS\end{tabular} & 5 & 2,123 & 3,673 & 8 & Medium \\
            \hline
            6 & \begin{tabular}[c]{@{}c@{}}Incident report: stolen AWS access keys\end{tabular} & 7 & 1,763 & 4,139 & 4 & Medium \\
            \hline
            7 & \begin{tabular}[c]{@{}c@{}}LUCR-3: Scattered Spider Getting SaaS-y \\ in the Cloud\end{tabular} & 2 & 3,413 & 4,392 & 13 & Low \\
            \hline
            8 & \begin{tabular}[c]{@{}c@{}}Ransomware in the cloud\end{tabular} & 7 & 4,111 & 6,430 & 11 & High \\
            \hline
            9 & \begin{tabular}[c]{@{}c@{}}SCARLETEEL: Operation leveraging Terraform, \\ Kubernetes, and AWS for data theft\end{tabular} & 11 & 3,254 & 7,748 & 20 & Medium \\
            \hline
            10 & \begin{tabular}[c]{@{}c@{}}Two real-life examples of why limiting \\ permissions works: Lessons from AWS CIRT\end{tabular} & 0 & 3,021 & 3,021 & 6 & Low \\
            \hline
            11 & \begin{tabular}[c]{@{}c@{}}Unmasking GUI-Vil: Financially Motivated \\ Cloud Threat Actor\end{tabular} & 5 & 7,021 & 9,606 & 15 & High \\
            \hline
            12 & \begin{tabular}[c]{@{}c@{}}When a Zero Day and Access Keys Collide in the Cloud: \\ Responding to the SugarCRM Zero-Day Vulnerability\end{tabular} & 6 & 4,718 & 5,783 & 15 & High \\
            \hline
        \end{tabular}
    \end{adjustbox}
    \caption{OSCTI sources used in our research.}
    \label{tab:osctis}
\end{table*}

\section{Running Example}\label{app:example}
This section provides a step-by-step demonstration of the \name framework in action.
The example uses an actual OSCTI source, specifically a Sysdig blog post titled \textit{“SCARLETEEL: Operation leveraging Terraform, Kubernetes, and AWS for data theft”}\footnote{\url{https://sysdig.com/blog/cloud-breach-terraform-data-theft/}}, which describes a cloud infrastructure exploit.
This demonstration focuses on specific paragraphs relevant to the \sig rule being generated, leading to the rule presented in Listing~\ref{lst:sigma_candidate_example}.

\begin{figure*}
  \centering
  \begin{subfigure}[t]{0.47915\textwidth}
    \includegraphics[width=\textwidth]{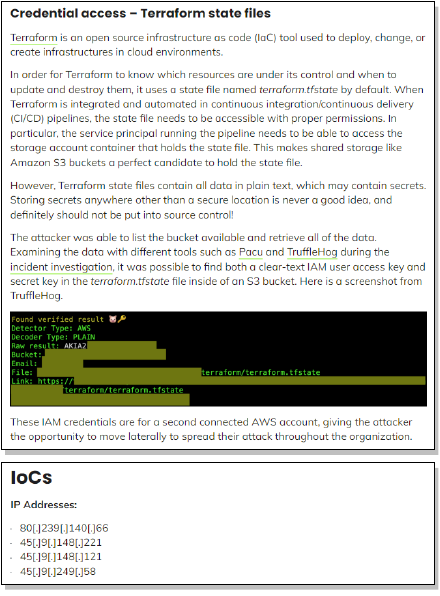}
    \caption{Screenshots of two paragraphs from the OSCTI}
    \label{fig:screenshots}
  \end{subfigure}
  \begin{subfigure}[t]{0.48\textwidth}
    \includegraphics[width=\textwidth]{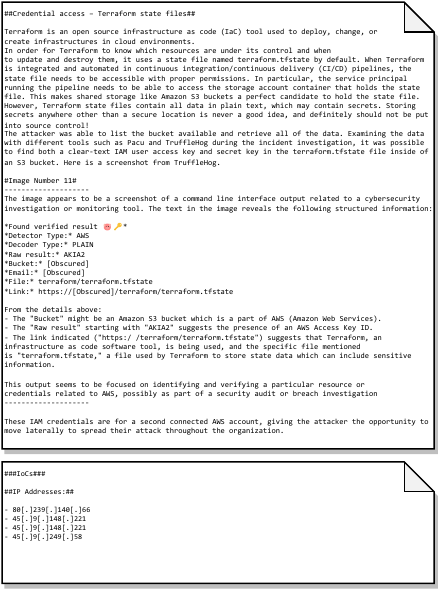}
    \caption{Corresponding preprocessed output}
    \label{fig:screenshots_analyses}
  \end{subfigure}
  \caption{OSCTI Preprocessing phase.}
  \label{fig:running_example_preprocessing}
\end{figure*}

\textbf{Step 1 (\textit{Preprocessing} Phase).} The initial phase involves preprocessing unstructured OSCTI data, which can be seen in Fig.~\ref{fig:screenshots}.
In this phase, the website content is downloaded and parsed by the \textit{Downloader and Parser} component, which converts HTML code into a structured format. 
The \textit{Image Analyzer} component processes the embedded images to extract relevant text.
This phase results in the formatted textual output shown in Fig.~\ref{fig:screenshots_analyses}.

\begin{figure*}
    \centering    \includegraphics{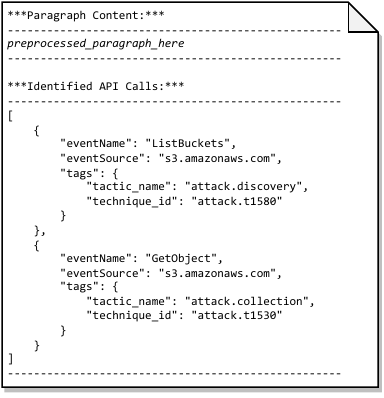}
    \caption{Formatted and enriched paragraph (input for the Rule Generator component).}
    \label{fig:rule_generator_input}
\end{figure*}

\textbf{Step 2 (\textit{Paragraph-Level Processing} Phase).} In this phase, each paragraph is processed to extract API calls and MITRE ATT\&CK TTPs using the \textit{API Call Extractor} and \textit{TTP Extractor} components.
These extractions are then attached to the formatted paragraph content to enrich it with additional context.
Fig.~\ref{fig:rule_generator_input} demonstrates how the API calls \textit{`ListBuckets'} and \textit{`GetObject'}, along with their event sources and TTPs, are added to the output in our example.
This enriched paragraph is then fed to the \textit{Rule Generator} component to generate an initial \sig rule, as shown in Listing ~\ref{lst:initial_sigma_candidate_example}.
\newline

\begin{lstlisting}[
    style=yaml,
    keywordstyle=\color{ProcessBlue},
    frame=single,
    keywords={title, id, status, description, references, author, date, modified, tags, logsource, product, service, detection, selection_event, eventSource, eventName, requestParameters, key, condition, falsepositives, level},
    literate={'}{{\textquotesingle}}1,
    caption=The initial generated Sigma rule.,
    label=lst:initial_sigma_candidate_example,
    captionpos=b
    ]
title: Access to Terraform File
description: Detects requests for terraform.tfstate file.
  This file contains sensitive infrastructure information
  and secrets, indicating potential compromise or
  unauthorized access.
references:
    - https://sysdig.com/blog/cloud-breach-terraform-data-
  theft/
    - https://docs.aws.amazon.com/AmazonS3/latest/API/
  API_GetObject.html
author: LLMCloudHunter
tags:
    - attack.collection
    - attack.t1530
logsource:
    product: aws
    service: cloudtrail
detection:
    selection_event:
        eventSource: s3.amazonaws.com
        eventName: GetObject
        requestParameters.key: terraform.tfstate
    condition: selection_event
falsepositives:
  - Automated CI/CD pipeline operations
  - DevOps engineers manually running Terraform commands
level: high
\end{lstlisting}

\textbf{Step 3 (\textit{OSCTI-Level Processing} Phase).} The \textit{Rule Optimizer} component refines the detection logic of each rule.
In our case, it finds no faults to fix and leaves the initial rule as it is.
The \textit{Set Refiner} removes duplicates in the overall set; here, there is no duplication of the \textit{`GetObject'} API call.
The \textit{IoC Enhancer} then uses the extracted IoCs in the IoC paragraph (Fig. ~\ref{fig:running_example_preprocessing}) to enhance the rule with suspicious IP addresses.
This results in the final \sig rule presented in Listing~\ref{lst:sigma_candidate_example}.

\section{Ablation study}\label{app:ablation}
In this section, we present the configurations used in the ablation study and our results. 
Table~\ref{tab:ablation_study_configurations} lists the different configurations, indicating which components were used in each variation.

\begin{table}[h]
    \centering
    \begin{adjustbox}{width=0.4\textwidth,center}
        \begin{tabular}{|c|c|c|c|c|c|c|c|c|c|c|}
            \hline
            \textbf{Name} & \multicolumn{10}{c|}{\textbf{Components Used}} \\
            \hline
            \noimagesmodel & A &   & C & D & E & F & G & H & I & J \\
            \noapimodel & A & B &   &   & E & F & G & H & I & J \\
            \nooptimizationmodel & A & B & C & D & E & F &   & H & I & J \\
            \textbf{\name} & A & B & C & D & E & F & G & H & I & J \\
            \hline
        \end{tabular}
    \end{adjustbox}
    \caption{Ablation study configurations.}
    \label{tab:ablation_study_configurations}
\end{table}

Table~\ref{tab:ablation_study_results} contains the results of the experiments performed in the ablation study, presenting the weighted average precision, recall, and F1 score for each variation and extraction type.

\begin{table*}[h]
    \centering
    \begin{adjustbox}{width=0.6\textwidth,center}
        \begin{tabular}{clcccc}
            \toprule
            \multirow{3}{*}{\textbf{Extraction}} & \textbf{Weighted} & \multirow{3}{*}{\textbf{\name}} & \multirow{3}{*}{\textbf{\noimagesmodel}} & \multirow{3}{*}{\textbf{\noapimodel}} & \multirow{3}{*}{\textbf{\nooptimizationmodel}} \\
                                                 & \textbf{Average} & & & & \\
                                                 & \textbf{Measure} & & & & \\
            \midrule
            \multirow{3}{*}{API Call} & Precision & 0.92 & 0.87 & 0.61 & 0.88 \\
            & Recall & 0.98 & 0.82 & 0.97 & 0.96 \\
            & F1 Score & 0.94 & 0.83 & 0.71 & 0.91 \\
            \midrule
            \multirow{3}{*}{Tactic} & Precision & 0.73 & 0.71 & 0.47 & 0.69 \\
            & Recall & 0.76 & 0.73 & 0.71 & 0.74 \\
            & F1 Score & 0.74 & 0.71 & 0.55 & 0.70 \\
            \midrule
            \multirow{3}{*}{Technique} & Precision & 0.62 & 0.58 & 0.24 & 0.57 \\
            & Recall & 0.75 & 0.59 & 0.27 & 0.62 \\
            & F1 Score & 0.68 & 0.57 & 0.24 & 0.58 \\
            \midrule
            \multirow{3}{*}{Sub-technique} & Precision & 0.65 & 0.50 & 0.29 & 0.63 \\
            & Recall & 0.71 & 0.53 & 0.24 & 0.64 \\
            & F1 Score & 0.67 & 0.50 & 0.25 & 0.62 \\
            \midrule
            \multirow{3}{*}{IoC} & Precision & 0.99 & 0.93 & 0.96 & 0.96 \\
            & Recall & 0.98 & 0.90 & 0.98 & 0.97 \\
            & F1 Score & 0.98 & 0.90 & 0.97 & 0.96 \\
            \midrule
            \multirow{3}{*}{Other} & Precision & 0.88 & 0.74 & 0.37 & 0.75 \\
            & Recall & 0.82 & 0.51 & 0.47 & 0.67 \\
            & F1 Score & 0.82 & 0.56 & 0.39 & 0.69 \\
            \midrule
            \multirow{3}{*}{Detection Entity $\leftrightarrow$ Sigma Field} & Precision & 0.87 & 0.83 & 0.58 & 0.83 \\
            & Recall & 0.90 & 0.72 & 0.79 & 0.84 \\
            & F1 Score & 0.88 & 0.75 & 0.65 & 0.83 \\
            \midrule
            \multirow{3}{*}{API Call $\leftrightarrow$ Tactic} & Precision & 0.77 & 0.57 & 0.37 & 0.72 \\
            & Recall & 0.80 & 0.41 & 0.54 & 0.73 \\
            & F1 Score & 0.78 & 0.46 & 0.42 & 0.72 \\
            \midrule
            \multirow{3}{*}{API Call $\leftrightarrow$ Technique} & Precision & 0.56 & 0.44 & 0.10 & 0.46 \\
            & Recall & 0.64 & 0.42 & 0.13 & 0.49 \\
            & F1 Score & 0.59 & 0.42 & 0.11 & 0.47 \\
            \midrule
            \multirow{3}{*}{API Call $\leftrightarrow$ Sub-technique} & Precision & 0.56 & 0.36 & 0.09 & 0.53 \\
            & Recall & 0.58 & 0.35 & 0.11 & 0.54 \\
            & F1 Score & 0.56 & 0.35 & 0.09 & 0.51 \\
            \midrule
            \multirow{3}{*}{API Call $\leftrightarrow$ IoC} & Precision & 0.92 & 0.94 & 0.70 & 0.88 \\
            & Recall & 0.91 & 0.75 & 0.93  & 0.87 \\
            & F1 Score & 0.92 & 0.83 & 0.78 & 0.87 \\
            \midrule
            \multirow{3}{*}{API Call $\leftrightarrow$ Other} & Precision & 0.84 & 0.74 & 0.41 & 0.72 \\
            & Recall & 0.85 & 0.61 & 0.41 & 0.76 \\
            & F1 Score & 0.84 & 0.65 & 0.39 & 0.72 \\
            \bottomrule
        \end{tabular}
    \end{adjustbox}
    \caption{Ablation study results.}
    \label{tab:ablation_study_results}
\end{table*}

\section{Component Configurations}\label{app:component_configurations}
In this section, we provide details on the configuration of each component in the \name framework.
For each component, Table~\ref{tab:component_configurations} lists the purpose, techniques, and parameters.

\begin{table*}[h]
    \centering
    \begin{adjustbox}{width=0.7\textwidth,center}
        \begin{tabular}{|c|c|c|c|c|c|c|}
            \hline
            \multirow{2}{*}{\textbf{Component}} & \multirow{2}{*}{\textbf{Purpose}} & \textbf{LLM}         & \textbf{Structured} & \textbf{Leverage} & \multirow{2}{*}{\textbf{Temperature}} & \textbf{Parallel} \\
                                                &                                   & \textbf{Utilization} & \textbf{Response}   & \textbf{Few-Shot} &                                       & \textbf{Requests} \\
            \hline
            \textbf{A}                          & HTML downloading and parsing      &                      &                     &                   &                                       &                   \\
            \hline
            \textbf{B}                          & Image processing                  & \checkmark           &                     &                   & 1                                     & \checkmark        \\
            \hline
            \multirow{2}{*}{\textbf{C}}         & Explicit API call extracting      & \checkmark           & \checkmark          &                   & 0                                     & \checkmark        \\
            \cline{2-7}
                                                & Implicit API call extracting      & \checkmark           & \checkmark          & \checkmark        & 0.9                                   & \checkmark        \\
            \hline
            \textbf{D}                          & TTPs extracting                   & \checkmark           & \checkmark          & \checkmark        & 0.5                                   & \checkmark        \\
            \hline
            \textbf{E}                          & Initial candidates generating     & \checkmark           & \checkmark          &                   & 0.7                                   & \checkmark        \\
            \hline
            \textbf{F}                          & Candidates validating             &                      &                     &                   &                                       & \checkmark        \\
            \hline
            \textbf{G}                          & Candidates optimizing             & \checkmark           & \checkmark          & \checkmark        & 0.5                                   & \checkmark        \\
            \hline
            \multirow{3}{*}{\textbf{H}}         & Duplicates extracting             &                      &                     &                   &                                       &                   \\
            \cline{2-7}                         & Candidate selecting               & \checkmark           & \checkmark          & \checkmark        & 0.5                                   &                   \\
            \cline{2-7}                         & API call removing                 & \checkmark           & \checkmark          & \checkmark        & 0.5                                   &                   \\
            \hline
            \textbf{I}                          & IoC extracting                    & \checkmark           & \checkmark          &                   & 0.5                                   &                   \\
            \hline
            \textbf{J}                          & Candidates IoC-enhancing          &                      &                     &                   &                                       &                   \\
            \hline
        \end{tabular}
    \end{adjustbox}
    \caption{Configuration of \name's components.}
    \label{tab:component_configurations}
\end{table*}

\end{document}